\documentclass[%
twocolumn,
nofootinbib,
aps,
floatfix,longbibliography
]{revtex4-2}

\usepackage{envLatex}

\newcommand{\myscale}{0.7}
\RequirePackage[colorlinks=true]{hyperref}

\hypersetup{colorlinks,breaklinks,
            pdfborder = {0 0 0},
            citecolor=[rgb]{0.0,0.5,0.5},
            urlcolor=[rgb]{0.0,0.5,0.5},
            linkcolor=[rgb]{0.0,0.5,0.5},
            citebordercolor=[rgb]{0.95,0.95,0.95},
            urlbordercolor=[rgb]{0.95,0.95,0.95},
            linkbordercolor=[rgb]{0.95,0.95,0.95},
           }

\begin{document}

\preprint{APS/123-QED}

\title{Quantum Advantage in Information Retrieval}

\author{Pierre-Emmanuel Emeriau}
\affiliation{
 Laboratoire d'Informatique de Paris 6
 \\ CNRS and Sorbonne Universit\'e
}
\email{pierre-emmanuel.emeriau@lip6.fr}

\author{Mark Howard}
\affiliation{
School of Mathematics, Statistics \& Applied Mathematics \\
National University of Ireland, Galway
}
\email{mark@markhoward.info}

\author{Shane Mansfield}
\affiliation{
Quandela \\
10 Boulevard Thomas Gobert, 91120, Palaiseau, France
}
\email{shane.mansfield@quandela.com}


\begin{abstract}
Random access codes have provided many examples of quantum advantage in communication,
but concern only one kind of information retrieval task.
We introduce a related task -- the Torpedo Game --
and show that it admits greater quantum advantage than the comparable random access code.
Perfect quantum strategies involving prepare-and-measure protocols with experimentally accessible three-level systems
emerge via analysis in terms of the discrete Wigner function.
The example is leveraged to an operational advantage in a pacifist version of the strategy game \textit{Battleship}.
We pinpoint a characteristic of quantum systems that enables quantum advantage  
in any bounded-memory information retrieval task.
While preparation contextuality has previously been linked to advantages in random access coding
we focus here on a different characteristic called
sequential contextuality.
It is shown not only to be necessary and sufficient for quantum advantage,
but also to quantify the degree of advantage.
Our perfect qutrit strategy for the Torpedo Game entails the strongest type of inconsistency with non-contextual hidden variables, revealing logical paradoxes with respect to those assumptions.
\end{abstract}


\maketitle


\onecolumngrid

\begin{figure}[h!]
        \centering
        \includegraphics[scale=1.5]{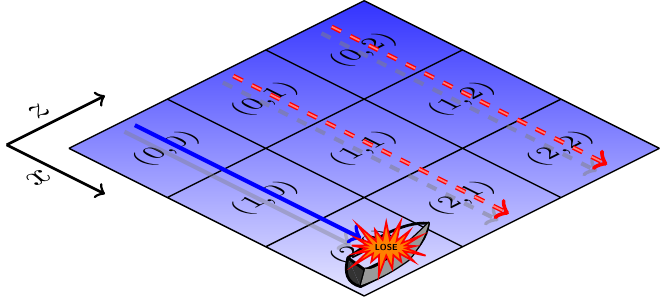}
        \caption{The Torpedo Game is a pacificist alternative to \textit{Battleship} where the aim is to avoid sinking Alice's ship, depicted here in dimension $3$.}
        \label{fig:TorpedoCartoon}
\end{figure}

\twocolumngrid

    \section*{Introduction}
        Random access coding involves the encoding of a random input string into a
shorter message string. The encoding should be such that any element of the
input string can be retrieved with high probability from the message string.
Such tasks have long been studied as examples in which the communication of
quantum information can provide advantage, \ie enhanced performance,
over classical information, e.g.\
\cite{ambainis1999dense,pawlowski2009information,spekkens2009preparation,grudka2014popescu,tavakoli2015quantum,chailloux2016optimal,aguilar2018connections,farkas2019self}.

However, random access coding concerns only one kind of information retrieval. In
this work we introduce another such task -- the Torpedo Game.
It is similar to random access coding, but with additional requirements involving the retrieval of {relative} information about elements
of the input string.
Taking a geometric perspective it may also be viewed as a pacifist
version of the popular strategy game \stress{Battleship} (see \refig{TorpedoCartoon}).
Quantum strategies can be implemented in prepare-and-measure scenarios, and
outperform classical strategies for the Torpedo Games with bit and trit inputs.
In particular, quantum perfect strategies exist in the trit case and provide a
greater quantum advantage than for the comparable random access coding task
\cite{tavakoli2015quantum}.

Optimal quantum strategies emerge from an analysis in terms of the discrete
Wigner function. Wigner negativity is a signature of non-classicality in
quantum systems that is related to contextuality and that has been widely studied
as a resource for quantum speed-up and advantage
\cite{galvao2005discrete,veitch2012negative,howard2014contextuality,delfosse2015wigner,bermejo2017contextuality,pashayan2015estimating,catani2018state,saha2019state}.
Knowing which characteristic lies at the source of better-than-classical performances can both allow for comparison of quantum systems in terms of their utility,
and offer a heuristic for generating further examples of quantum-enhanced performance.
Our optimal quantum strategies are indeed Wigner negative, with perfect quantum strategies derived from maximum Wigner negativity.
Yet while negativity is necessary for advantage in the Torpedo game, it is not sufficient.

To more precisely pinpoint the source of quantum advantage we must
look further.
One candidate would be preparation contextuality \cite{spekkens2005contextuality}, another signature of non-classicality that has been linked to QRACs in numerous studies \cite{spekkens2009preparation,chailloux2016optimal,ambainis2019parity}.
It has been shown to be necessary for advantage in a restricted class of random access codes subject to an obliviousness constraint \cite{hameedi2017communication,saha2019state}.

In this work, however, we focus on a different characteristic called sequential contextuality \cite{mansfield2018quantum}.
It indicates the absence of a hidden variable model respecting the sequential structure of a given protocol.
Subject to an assumption of bounded-memory,
we find that this characteristic is necessary and sufficient for quantum advantage, not just in random access coding but in any information retrieval task.
Moreover, we show that it quantifies the degree of advantage that can be achieved.

Contextuality can exhibit itself at the level of probability distributions, e.g.\ quantum violations of the CHSH inequality \cite{chsh}.
But in some cases it can also manifest at the level of the supports
of these distributions.
In other words contextuality can be inferred by a series of logical deductions about which events are possible or not, e.g.\ Hardy's paradox \cite{hardy92,hardy93}.
This situation has come to be known as logical contextuality \cite{ab}.
In the most extreme cases, known as strong contextuality \cite{ab}, every possible event triggers such a paradox \cite{mansfield2017consequences}, e.g.\ Popescu-Rohrlich box violations of the CHSH inequality \cite{popescu1994quantum} (see also \cite{abramsky_et_al:LIPIcs:2015:5416,de2018logical}).
Note that empirical data is always expressible as a convex mixture of classical and strongly contextual data \cite{ab,abramsky2017contextual}.
For qutrits the quantum perfect strategies we introduce display analogous contextuality, and hence paradoxes, of this strongest form in a prepare-and measure scenario.


\refsec{irtasks} gives an overview of information retrieval tasks including random access coding and the Torpedo Game. \refsec{DWF} provides background on discrete Wigner functions. \refsec{optima} deals with optimal classical and quantum strategies for the Torpedo Game. Finally, \refsec{context} establishes the relationship between sequential contextuality and quantum advantage in bounded-memory information retrieval tasks.

    \section{Information Retrieval Tasks}
        \label{sec:irtasks}
\subsection{Random Access Codes}
\label{subsec:RAC}

An $(n,m)_2$ {Random Access Code (RAC)}, sometimes denoted $n \rightarrow m$, is a communication task in which one aims to encode information about a random $n$-bit input string into an $m$-bit message where $m<n$, in such a way that any one of the input bits may be retrieved from the message with high probability.
An $(n,m)_2$ {Quantum Random Access Code (QRAC)} instead encodes the input into an $m$-qubit (quantum) message state.

Such tasks may be considered as two-party cooperative games in which the first party, Alice, receives a random input string from a referee.
She encodes information about this in a message that is communicated to the second party, Bob.
The referee then asks Bob to retrieve the value of the bit at a randomly chosen position in the input string.
We will assume that the referee's choices are made uniformly at random.

For instance, for the $(2,1)_2$ RAC game \cite{ambainis1999dense} an optimal classical strategy is for Alice to directly communicate one of the input bits to Bob.
If asked for this bit, Bob can always return the correct answer, otherwise Bob guesses and will provide the correct answer with probability $\frac{1}{2}$.
Thus the game has a classical value of $$\theta^C_{2\rightarrow 1} = \frac{1}{2} \left( 1 + \frac{1}{2} \right) = \frac{3}{4}\, .$$
Quantum strategies can outperform this classical bound.
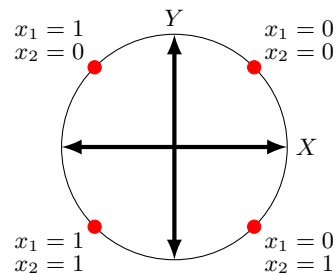
\begin{figure}[htbp]
    \centering
    \begin{tikzpicture}[scale=.5]

\draw (0,0) circle (3cm);
\draw [<->,ultra thick] (-3,0) -- (3,0);
\draw [<->,ultra thick] (0,-3) -- (0,3);

\node [right] (X) at (3,0) {$X$};
\node [above] (Y) at (0,3) {$Y$};

\node (i00) at (2.121,2.121) {};
\node (texte000) at ($(i00)+(1.2,1)$) {$x_1=0$};
\node (texte001) at ($(i00)+(1.2,.4)$) {$x_2=0$};
\draw [fill=red,red] (i00) circle (5pt);

\node (i01) at (2.121,-2.121) {};
\node (texte010) at ($(i01)+(1.2,-.4)$) {$x_1=0$};
\node (texte011) at ($(i01)+(1.2,-1)$) {$x_2=1$};
\draw [fill=red,red] (i01) circle (5pt);

\node (i10) at (-2.121,2.121) {};
\node (texte100) at ($(i10)+(-1.2,1)$) {$x_1=1$};
\node (texte101) at ($(i10)+(-1.2,.4)$) {$x_2=0$};
\draw [fill=red,red] (i10) circle (5pt);

\node (i11) at (-2.121,-2.121) {};
\node (texte110) at ($(i11)+(-1.2,-.4)$) {$x_1=1$};
\node (texte111) at ($(i11)+(-1.2,-1)$) {$x_2=1$};
\draw [fill=red,red] (i11) circle (5pt);

\end{tikzpicture}
    \caption{The four red dots correspond to the four states $\ket{\psi_{x_1,x_2}}$ defined in \refequ{QRAC21_optimalstate} depicted as points on the equator of the Bloch sphere.}
    \label{fig:QRAC21}
\end{figure}

An optimal quantum strategy is for Alice to communicate the qubit state
\begin{equation}
    \ket{\psi_{x_1,x_2}} = \frac{1}{\sqrt{2}} \left( \ket{0} +  \frac{1}{\sqrt{2}} \left(  (-1)^{x_1} + (-1)^{x_2} i \right) \ket{1} \right)
    \label{eq:QRAC21_optimalstate}
\end{equation}
where $(x,z)$ is the input bit-string she has received.
Bob measures in the $X$ basis when asked for $x_1$ and in the $Y$ basis when asked for $x_2$ (see \refig{QRAC21}).
If he obtains the $+1$ eigenvalue he returns the value $1$ and if he obtains the $-1$ eigenvalue he returns $0$.
This yields a quantum value for the game of $$\theta^Q_{2\rightarrow 1} = \cos^2\left(\frac{\pi}{8}\right) \approx 0.85 \, .$$

\subsection{General Information Retrieval Tasks}

One may also consider more general communication scenarios.
In an $(n,m)_d$ {communication scenario} the input is a random string of $n$ $d$its and the message is a string of $m$ (qu)$d$its, for $d \geq 2$.
(Q)RAC tasks have previously been considered in such scenarios, e.g.\ in \cite{tavakoli2015quantum,casaccino2008extrema}.

However, we also wish to accommodate for a much wider range of information retrieval tasks.
An {information retrieval task} in an $(n,m)_d$ communication scenario is specified by a tuple $\langle Q, \{w_q\}_{q \in Q} \rangle$.
\begin{itemize}
\item $Q$ is a finite set of \stress{questions}.
\item The $w_q : \mathbb{Z}_d^n \rightarrow \mathbb{Z}_d$ are {winning relations}, which pick out the good answers to question $q$ given an input string in $\mathbb{Z}_d^n$. Note that there may be more than one good answer, or none. It is assumed that inputs and outputs are endowed with the structure of the commutative ring $\mathbb{Z}_d$.
\end{itemize}

Standard $(n,m)_d$ (Q)RACs are recovered when the questions ask precisely for the respective input dits.
In that case the winning relations $w_{i}$ reduces simply to projectors onto the respective dits of the input string.
However, other interesting tasks arise when the questions also concern relative information about the input string, in the form of parities or linear combinations modulo $d$ of the input dits.
A similar generalisation for $d=2$, using functions rather than relations, has been independently proposed in \cite{doriguello2020quantum}.

\subsection{The Torpedo Game}
\label{sec:torpedo_game}

\noindent
Of particular interest in the present work is an information retrieval task for $(2,1)_d$ communication scenarios.
We take the game perspective and refer to the task as the dimension $d$ {Torpedo Game} (see \refig{TorpedoCartoon} in dimension 3).
Let $x$ and $z$ be the two input dits.
There are $d+1$ questions 
$Q = \{ \infty, 0, 1, \dots, d-1 \}$. The labelling comes from a geometric interpretation to be elaborated upon shortly.
Winning relations for the Torpedo Game are given by
\begin{align}
\begin{split}
w_\infty (x,z) &= \{ a \in \mathbb{Z}_d \mid a \neq x \} \\
w_0 (x,z) &= \{ a \in \mathbb{Z}_d \mid a \neq -z \} \\
w_1 (x,z) &= \{ a \in \mathbb{Z}_d \mid a \neq x-z \} \\
w_2 (x,z) &= \{ a \in \mathbb{Z}_d \mid a \neq 2x-z \} \\
& \vdots \\
w_{d-1} (x,z) &= \{ a \in \mathbb{Z}_d \mid a \neq (d-1)x-z \} \, .
\end{split}
\label{eq:winning_conditions}
\end{align}
All arithmetic is modulo $d$.

For $d=2$, the Torpedo Game is equivalent to a $(2,1)_2$ (Q)RAC, but with an additional question.
Bob may be asked to retrieve either one the individual input dits, or to retrieve relative information about them in the form of their parity $x \oplus z$. We note that dimension 2 is the only case where the winning relations are actually functions (there is only one good answer per question).

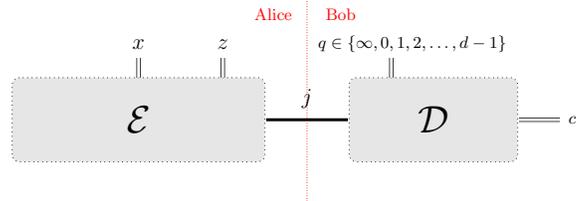
\begin{figure}[htbp]
    \centering
    \scalebox{\myscale}{    \usetikzlibrary{arrows,3d,fadings,shapes,calc,decorations.pathreplacing,decorations.markings,positioning}
\def\tkzscl{0.8}

\begin{tikzpicture}[scale=\tkzscl]

\node (up) at (0,1) {};
\node (down) at (0,-1) {};
\node (left) at (-1,0) {};
\node (right) at (1,0) {};

\node (a1) at (0,0) {};
\node (a2) at (6,2) {};
\node [inner sep=.0pt] (a3) at (6,1) {};

\node (b1) at (8,0) {};
\node (b2) at (12,2) {};
\node [inner sep=.0pt] (b3) at (8,1) {};

\draw[rounded corners,dotted,fill=gray!20] (a1) rectangle (a2);
\node (am) at (2,1) {};
\node (E) at ($(am)+(right)$) {\huge $\mathcal{E}$};
\node [inner sep=0 pt] (d1) at ($(am)+1.5*(up)+(right)$) {};
\node [inner sep=0 pt] (d2) at ($(am)+1.5*(up)+3*(right)$) {};
\draw [double distance=1.5pt] (d1) to ($(d1)+.5*(down)$);
\draw [double distance=1.5pt] (d2) to ($(d2)+.5*(down)$);
\node (d1texte) at ($(d1)+.3*(up)$) {\large $x$};
\node (d2texte) at ($(d2)+.3*(up)$) {\large $z$};

\draw[rounded corners,dotted,fill=gray!20,dotted] (b1) rectangle (b2);
\node (bm) at (10,1) {\huge $\Dc$};
\node [inner sep=0 pt] (j) at ($(bm)+1.5*(up)+(left)$) {};
\draw [double distance=1.5pt] (j) to ($(j)+.5*(down)$);
\node (jtexte) at ($(j)+.3*(up)+.5*(right)$) {$q \in \{\infty,0,1,2, \dots, d-1 \}$};
\node [inner sep=0 pt] (o) at ($(bm)+2*(right)$) {};
\draw [double distance=1.5pt] (o) to ($(o)+(right)$);
\node (otexte) at ($(o)+1.3*(right)$) {$c$};

\draw[ultra thick] (a3) to (b3); 
\draw[densely dotted, red] ($.5*(a2)+.5*(b1)+3*(up)$) to ($.5*(a2)+.5*(b1)+2*(down)$);
\node (Alice) at ($.5*(a3)+.5*(b3)+2.5*(up)+.8*(left)$) {\textcolor{red}{Alice}};
\node (Bob) at ($.5*(a3)+.5*(b3)+2.5*(up)+.8*(right)$) {\textcolor{red}{Bob}};
\node (psi) at ($.5*(a2)+.5*(b1)+.5*(up)$) {\large $j$};

\end{tikzpicture}}
    \caption{Prepare-and-measure protocol for the Torpedo Game: Alice receives dits $x$ and $z$ and sends a single message (qu)dit $j$ via the encoding $\Ec$. Bob is asked a question ${q \in \{ \infty,0,\dots,d-1 \}}$, performs decoding $\Dc$, and outputs $c$ which should satisfy the winning conditions given by $w_q (x,z)$ with high probability.} 
    \label{fig:QRAC_pres}
\end{figure}

The Torpedo Game may be framed as cooperative, pacifist alternative to the popular game \stress{Battleship},
in which Alice and Bob, finding themselves on opposing sides in a context of naval warfare, wish to subvert the conflict and cooperate to avoid casualities while not directly disobeying orders.

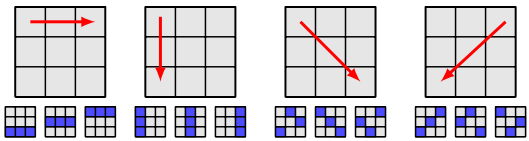
\begin{figure}[htbp]
    \centering
    \scalebox{\myscale}{\begin{tikzpicture}[scale=.19]

\draw [thick, fill=gray!20,step=3] (0,0) grid (9,9) rectangle (0,0);
\draw [->,ultra thick, red] (1.5,7.5) -- (8,7.5);

\draw [fill=gray!20] (3,-4) rectangle (6,-1);
\draw [fill=blue!70] (3,-3) rectangle (6,-2);
\draw [thick, fill=gray!20] (3,-4) grid (6,-1);

\draw [fill=gray!20] (-1,-4) rectangle (2,-1);
\draw [fill=blue!70] (-1,-4) rectangle (2,-3);
\draw [thick, fill=gray!20] (-1,-4) grid (2,-1);

\draw [fill=gray!20] (7,-4) rectangle (10,-1);
\draw [fill=blue!70] (7,-2) rectangle (10,-1);
\draw [thick, fill=gray!20] (7,-4) grid (10,-1);

\draw [thick, fill=gray!20,step=3,xshift=1cm] (12,0) grid (21,9) rectangle (12,0);
\draw [ultra thick, red, <-] (14.5,1.5) to (14.5,8);

\draw [fill=gray!20] (12,-4) rectangle (15,-1);
\draw [fill=blue!70] (12,-4) rectangle (13,-1);
\draw [thick, fill=gray!20] (12,-4) grid (15,-1);

\draw [fill=gray!20] (16,-4) rectangle (19,-1);
\draw [fill=blue!70] (17,-4) rectangle (18,-1);
\draw [thick, fill=gray!20] (16,-4) grid (19,-1);

\draw [fill=gray!20] (20,-4) rectangle (23,-1);
\draw [fill=blue!70] (22,-4) rectangle (23,-1);
\draw [thick, fill=gray!20] (20,-4) grid (23,-1);

\draw [thick, fill=gray!20,step=3] (27,0) grid (36,9) rectangle (27,0);
\draw [ultra thick, red, <-] (34.5,1.5) to (28.5,7.5);

\draw [fill=gray!20] (26,-4) rectangle (29,-1);
\draw [fill=blue!70] (26,-4) rectangle (27,-3);
\draw [fill=blue!70] (27,-2) rectangle (28,-1);
\draw [fill=blue!70] (28,-3) rectangle (29,-2);
\draw [thick, fill=gray!20] (26,-4) grid (29,-1);

\draw [fill=gray!20] (30,-4) rectangle (33,-1);
\draw [fill=blue!70] (30,-2) rectangle (31,-1);
\draw [fill=blue!70] (31,-3) rectangle (32,-2);
\draw [fill=blue!70] (32,-4) rectangle (33,-3);
\draw [thick, fill=gray!20] (30,-4) grid (33,-1);

\draw [fill=gray!20] (34,-4) rectangle (37,-1);
\draw [fill=blue!70] (34,-3) rectangle (35,-2);
\draw [fill=blue!70] (35,-4) rectangle (36,-3);
\draw [fill=blue!70] (36,-2) rectangle (37,-1);
\draw [thick, fill=gray!20] (34,-4) grid (37,-1);

\draw [thick, fill=gray!20,step=3,xshift=2cm] (39,0) grid (48,9) rectangle (39,0);
\draw [ultra thick, red, <-] (42.5,1.5) to (49,7.5);

\draw [fill=gray!20] (40,-4) rectangle (43,-1);
\draw [fill=blue!70] (40,-4) rectangle (41,-3);
\draw [fill=blue!70] (41,-3) rectangle (42,-2);
\draw [fill=blue!70] (42,-2) rectangle (43,-1);
\draw [thick, fill=gray!20] (40,-4) grid (43,-1);

\draw [fill=gray!20] (44,-4) rectangle (47,-1);
\draw [fill=blue!70] (44,-3) rectangle (45,-2);
\draw [fill=blue!70] (45,-2) rectangle (46,-1);
\draw [fill=blue!70] (46,-4) rectangle (47,-3);
\draw [thick, fill=gray!20] (44,-4) grid (47,-1);

\draw [fill=gray!20] (48,-4) rectangle (51,-1);
\draw [fill=blue!70] (48,-2) rectangle (49,-1);
\draw [fill=blue!70] (49,-4) rectangle (50,-3);
\draw [fill=blue!70] (50,-3) rectangle (51,-2);
\draw [thick, fill=gray!20] (48,-4) grid (51,-1);

\end{tikzpicture}}
    \caption{The red arrows depict the directions or slopes ($\infty, 0, 1, 2$, respectively) along which Bob may be asked to shoot in the $d=3$ Torpedo Game. For each direction, Bob has three possibilities, depicted by the blue lines. In the affine plane of order 3, each of these groups of three blue cells forms a line.} 
    \label{fig:d3_torpedo_firing}
\end{figure}

We take the input dits received by Alice as designating the coordinates in which she is ordered by her commander to position her one-cell ship on the affine plane of order $d$.
We may think of the affine plane as a toric $d \times d$ grid, with $x$ designating the row and $z$ the column.
E.g.\ in \refig{d3_torpedo_firing} we identify the top edge with the bottom edge and the left edge with the right edge.

Bob is a naval officer on the opposing side who is ordered by his commander to shoot a torpedo along a line of the grid with slope specified by $q \in Q$.
The $\infty$ question requires Bob to shoot along some row, and the $0$ question requires Bob to shoot along some column, etc.
However, Bob retains the freedom to choose which row, or column, or diagonal of given slope, as the case may be.
In other terms, upon receiving $q$ Bob must shoot along a lines $q x - z = c$ (if $q \neq \infty$) or $x = c$ (if $q = \infty$) but is free to choose the constant $c$.

Alice and Bob wish to coordinate a strategy for avoiding casualities, while still obeying their explicit orders.
Alice may communicate a single (qu)dit to Bob
-- greater communication may risk revealing her position should it be intercepted --
and based on this Bob must choose his $c$ in such a way that he avoids Alice's ship.

    \section{The Discrete Wigner Function}
        \label{sec:DWF}
        
It is possible to represent finite-dimensional quantum states as quasi-probability distributions over a phase space of discrete points.
Wootters \cite{gibbons2004discrete,wootters1987wigner}
introduced  a  method  of constructing discrete Wigner functions (DWF) based on finite fields, wherein vectors from a complete set of mutually unbiased bases in $\mathbb{C}^d$ are put in one-to-one correspondence with the lines of a finite affine plane of order $d$. This geometric picture of the DWF is useful for visualizing our Torpedo Game as exemplified in Fig.~\ref{fig:d3_torpedo_firing}, where each distinct orthonormal basis corresponds to a set of $d$ parallel (non-intersecting) lines.
Gross \cite{gross2006hudson}
singled out one particularly symmetric definition of DWF that obeys the discrete version of Hudson’s Theorem. This theorem says that an odd-dimensional pure state is non-negatively represented in the DWF if and only if it is a stabilizer state (defined below). The discrete  Hudson’s Theorem has remarkable implications, providing large classes of quantum circuit with a local hidden variable model that enables efficient simulation \cite{veitch2012negative,mari2012positive}. Clearly, negativity in this DWF is a necessary prerequisite for quantum speed-up. Howard \textit{et al.} \cite{howard2014contextuality} showed that this negativity actually corresponds to contextuality with respect to Pauli measurements, thereby establishing the operational utility of contextuality for the gate-based model of quantum computation (particularly in a fault-tolerant setting). The equivalence of Wigner negativity and contextuality was established by deriving a noncontextuality inequality using the graph-theoretic techniques of Cabello, Severini and Winter \cite{cabello2014graph} which extend Kochen-Specker type state-independent proofs to the state-dependent realm. This proof (and a subsequent alternate proof \cite{delfosse2017equivalence}) requires that, as well as the system displaying Wigner negativity, a second ancillary system must be present in order to have a sufficiently rich set of available measurements.

\subsection{DWF Formalism}
The discrete Wigner function is both foundationally interesting as well as practically relevant for fault-tolerant quantum computing via its link with so-called ``stabilizer states''. The qudit versions of the $X$ and $Z$ Pauli operators are
\begin{align*}
X\ket{k}&=\ket{k+1}\\
Z\ket{k}&=\omega^k\ket{k}
\end{align*}
where $\omega = \exp(2 \pi i /d)$ and arithmetic is modulo $d$. The qudit Pauli group has elements which are products of (powers of) these operators e.g. $X^xZ^z$ for $x,z \in \mathbb{Z}_d$. A unitary $U$ stabilizes a state $\ket{\psi}$ if $U\ket{\psi}=\ket{\psi}$. A stabilizer state is the unique $n$-qudit state stabilized by a subgroup of size $d^n$ of the Pauli group. Equivalently, stabilizer states may be understood as the image of computational basis states under the Clifford group, which is the set of unitaries that map the Pauli group to itself under conjugation.

For an arbitrary $d \times d$ Hermitian operator $Q$ of unit trace (typically a density matrix), its Wigner representation will consist of $d^2$ real quasi-probabilities $W_{x,z}$ for $x,z \in \mathbb{Z}_d$. In particular, the quasi-probability associated with the point $(x,z)\in\mathbb{Z}_d^2$ is given by
\begin{align*}
W_{x,z}=\frac{1}{d}\Tr(Q A_{x,z})
\end{align*}
where $A_{x,z}$ are the so-called phase point operators to be defined shortly. The unit trace of $Q$ will ensure that $\sum_{x,z} W_{x,z}=1$. Taking the magnitude $|W_{x,z}|$ of each quasi-probability will lead to $\sum_{x,z} |W_{x,z}|=1$ if and only if the quasi-probability distribution is actually a legitimate (non-negative) discrete probability distribution. In contrast, the presence of negative quasi-probabilities entails $\sum_{x,z} |W_{x,z}|>1$, and in fact the departure of $\sum_{x,z} |W_{x,z}|$ from unity is a sensible measure of ``how negative'' or ``how non-classical'' the DWF of an operator is \cite{veitch2012negative,veitch2014resource}.

When working with the DWF, it is convenient to use the Weyl-Heisenberg notation and phase convention for the qudit Pauli operators i.e.
\begin{align*}
	D_{x , z}=\omega^{2^{-1} x z} \sum_{k} \omega^{k z}|k+x\rangle\langle k|=\omega^{\frac{x z}{2}} X^{x} Z^{z}, 
\end{align*}
where they go by name displacement operators. The phase point operator at the origin of phase space $A_{0,0}$ is given by the simple expression
\begin{align*}
A_{0,0}=\sum_{j\in\mathbb{Z}_d} \ketbra{-j}{j},
\end{align*}
and the remainder are found by conjugation with displacement operators
\begin{align}
A_{x,z}=D_{x,z}A_{0,0}D_{x,z}^\dag\,. \label{eq:PPOdefn}
\end{align}

\subsection{DWF and Information Retrieval}

The eigenvectors of phase point operators are objects of interest. The maximizing eigenvectors of the phase point operators in \refequ{PPOdefn} (and additional ones from different choices of DWF) were used in Casaccino \textit{et al.} \cite{casaccino2008extrema} as the encoded messages of a $(d+1,1)_d$ QRAC. This is natural given the use of MUBs in constructing DWFs, and prominence of MUBs in the QRAC literature. If Alice receives input $\pmb{k}=(k_1,k_2,\ldots,k_{d+1})\in\mathbb{Z}_d^{d+1}$ that she encodes in  $\rho_{\pmb{k}}$ and transmits to Bob, then the average probability of success for the Casaccino \textit{et al.} QRAC is 
\begin{align}
		\frac{1}{(d+1)d^{d+1}}\sum_{\pmb{k}\in\mathbb{Z}_d^{d+1}}\Tr\left[\rho_{\pmb{k}}(\Pi_1^{k_1}+\ldots+ \Pi_{d+1}^{k_{d+1}})\right]\, \label{eq:GalvaoQRAC}
\end{align}
where $\Pi_q^{i}$ is the projector corresponding to dit value $i$ in Bob's $q$-th measurement setting. Since phase point operators are constructed using sums of projectors from MUBs i.e., $\Pi_1^{k_1}+\Pi_2^{k_2}+\ldots+ \Pi_{d+1}^{k_{d+1}}$, the use of a maximizing eigenvector of a phase point operator for $\rho_{\pmb{k}}$ is natural to maximize \refequ{GalvaoQRAC}.

In this work we instead make use of the \textit{minimizing} eigenvectors of phase point operators. The rationale for this is two-fold (i) these eigenvectors display remarkable geometric properties with respect to the measurements in (their constituent) mutually unbiased bases, and (ii) negativity (of a state in the DWF) is the hallmark of non-classicality which has already been identified with contextuality (with the already mentioned caveat that an additional ``spectator'' subsystem was required). These will be seen to lead to a perfect quantum strategy for the Torpedo Game.

As previously noted in \cite{gross2006hudson,van2011noise}, the eigenvectors of phase point operators \refequ{PPOdefn} are degenerate: a $+1$ eigenspace of dimension $\frac{d+1}{2}$ and a $-1$ eigenspace of dimension $\frac{d-1}{2}$.  Any state in the $-1$ eigenspace has an outcome that is forbidden \cite{van2011noise,bengtsson2012kochen} 
in each of a complete set of MUBs. For example, let $\ket{\psi_{0,0}}=(\ket{1}-\ket{d-1})/\sqrt{2}$ satisfying $A_{0,0}\ket{\psi_{0,0}}=-\ket{\psi_{0,0}}$. This state obeys $\Tr(\Pi_q^0 \ketbra{\psi_{0,0}}{\psi_{0,0}})=0$, where $\Pi^{0}_q$ is the projector on the $0$-th eigenvector in the $q$-th basis. More specifically, $\Pi^{0}_q$ is the projector corresponding to the $\omega^{0}=+1$ eigenvector of displacement operator $\left\{D_{0,1},D_{1,0},D_{1,1},\dots,D_{1,d-1}\right\}$. These displacement operators have eigenvectors leading to mutually unbiased measurement bases $q\in\{\infty,0,1,\dots,d-1\}$ respectively. The related states $\ket{\psi_{x,z}}=D_{x,z}\ket{\psi_{0,0}}$, which are eigenstates $A_{x,z}\ket{\psi_{x,z}}=-\ket{\psi_{x,z}}$,
obey
\begin{widetext}
\begin{equation}
	\Tr\left[\ketbra{\psi_{x,z}}{\psi_{x,z}}(\Pi_{\infty}^{x}+\Pi_0^{-z}+\Pi_1^{x-z}+\dots+\Pi_{d-1}^{(d-1)x-z})\right]=0\,. \label{eq:KeyFact}
\end{equation}
\end{widetext}
Equation \ref{eq:KeyFact} implies that probability of the relevant outcome (outcome $x$ in the first basis, $-z$ in the second basis, etc.) in each of the MUBs is zero: cf.\ Equation \ref{eq:winning_conditions}. The general expression for odd power-of-prime $d$ is proven in \cite{howard2015classical,appleby2008spectra}.

    \section{Optimal Strategies for the Torpedo Game}
        \label{sec:optima}
        
Here we gather the optimal classical, quantum and (in one case) post-quantum strategies for the Torpedo Game. We only focus on Torpedo games with power-of-prime dimension $d$ as we are able to provide perfect quantum strategies in these cases (for $d \geq 3$) due to the fact that there exist $d+1$ MUBs for those dimensions. The quantum case differs depending on whether we use a qubit or a qudit of odd prime power dimension. The classical optimum can only be established rigorously for small dimensions, owing to the proliferation of possible hidden variable assignments as the dimension increases. We obtain a quantum advantage for dimension 2 and 3. At the conclusion of this article we sketch a modified Torpedo Game that we believe may have a lower classical value whenever $d\geq 5$, thereby re-establishing a quantum advantage in those dimensions.

\subsection{Optimal Quantum and Post-Quantum Strategies}
\label{sec:qstrat}

\begin{figure}[ht]
\includegraphics[width=0.5\textwidth]{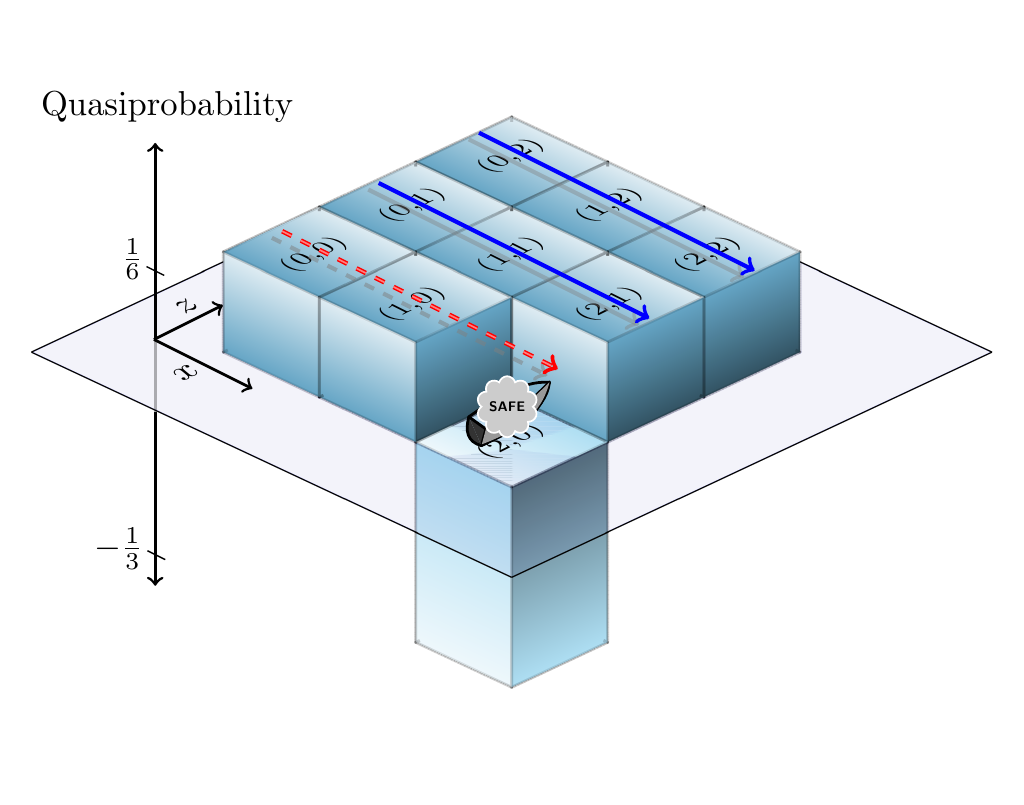}
        \caption{The perfect quantum strategy can be understood by plotting the Discrete Wigner Function of the message state sent by Alice. In the above qutrit case, Alice is in coordinate $(x,z)=(2,0)$, so she sends Bob the state $\ket{\psi_{2,0}}$ whose Wigner function is $-1/3$ at coordinate $(2,0)$ and $1/6$ otherwise. Bob measures this state along any of the four allowed directions, wherein the probability of each outcome is given by sum of quasiprobabilities along the corresponding line. Hence, the only outcomes with non-zero probability of occurring correspond to lines not passing through $(2,0)$. Whichever outcome Bob sees, he may fire his torpedo along the corresponding line, safe in the knowledge that it will not intersect Alice's ship. In this figure the solid blue lines correspond to the possible outcomes for the $q=0$ direction, but the same argument holds for all the other directions.
        }
        \label{fig:WigFigCartoon}
\end{figure}

\paragraph{Quantum Perfect Strategy for Odd Power-of-Prime Dimension.} 
From \refequ{KeyFact} it follows that there is a perfect quantum strategy for the dimension $d$ Torpedo game for any for odd power-of-prime $d$:
\begin{enumerate}
    \item Upon receiving dits $x$ and $z$ Alice sends the following state to Bob:
        \begin{equation}
            \ket{\psi_{x,z}} = D_{x,z} \ket{\psi_{0,0}} = D_{x,z} \left( \sqrt{2}^{-1}(\ket{1} - \ket{-1}) \right). \label{eq:PerfectState}
        \end{equation}
    \item Bob receives $\ket{\psi_{x,z}}$ and is asked a question $q \in \{\infty,0, \dots, d-1 \}$.
    He measures the state in the MUB corresponding to $q$ and outputs the dit corresponding to the measurement outcome.
\end{enumerate}
This quantum strategy wins the Torpedo Game deterministically, \ie with probability $1$. In \refig{WigFigCartoon} we provide geometric intuition for why this strategy is perfect.

\paragraph{Optimal Quantum Strategy for Qubits.}
An analogous strategy to the qudit case can be employed for the qubit Torpedo Game, using message states $\ket{\psi_{x,z}} = X^xZ^z \ket{\psi_{0,0}}$ where $X,Y$ and $Z$ are the usual qubit Pauli spin matrices and $\ketbra{\psi_{0,0}}{\psi_{0,0}}=\frac{1}{2}\left(\mathbb{I}-(X+Y+Z)/\sqrt{3}\right)$. For $d=2$, while this does not constitute a perfect strategy it still achieves an advantage over classical strategies.
In fact, it turns out to be an optimal strategy: this strategy achieves a winning probability of approximately 0.79 and we show that this is optimal. First we can leverage the fact that the $(3,1)_2$ (Q)RAC attributed to Isaac Chuang is at least as hard to win as the Torpedo Game because for the last question, instead of asking to retrieve the parity of the input bits, the $(3,1)_2$ (Q)RAC ask to retrieve a third independent bit. Thus we get a lower bound of $\frac{1}{2}\left(1+\frac{1}{\sqrt{3}}\right) \approx 0.79$ on the optimal quantum value. To obtain a matching upper bound, we implemented numerically the NPA hierarchy \cite{navascues2008convergent} which is a hierarchy of semi-definite programs converging from the exterior to the correlations arising from quantum systems. Because the message sent from Alice to Bob is of finite dimension we relied mostly on \cite{navascues2015characterizing} which allows to characterise correlations arising from finite-dimensional quantum systems. We found a matching upper bound proving that indeed $\theta_{d=2}^Q \approx 0.79$.

\paragraph{Perfect Post-quantum Strategy for Qubits.}

\begin{figure}[htbp]
        \centering
        \includegraphics[scale=0.2]{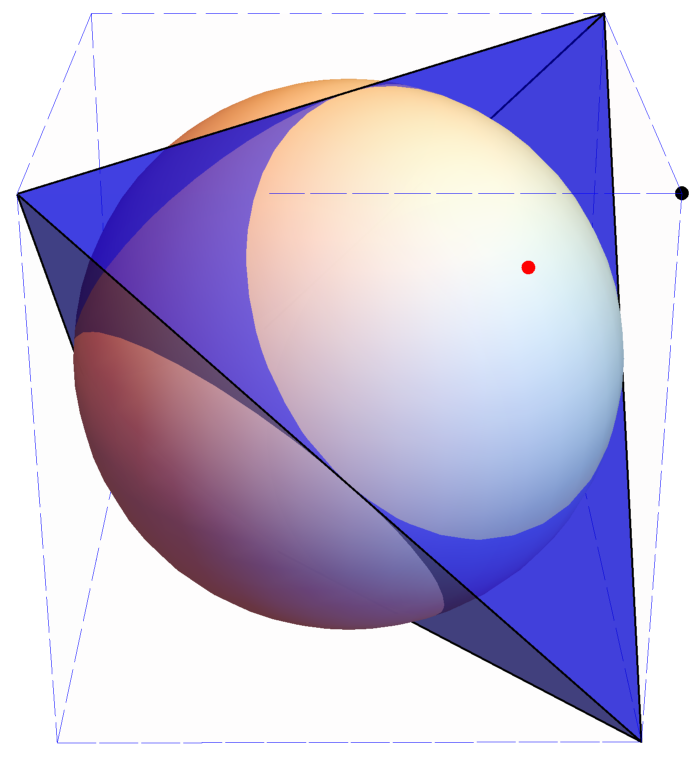}
        \caption{The qubit version of the Torpedo Game has a perfect strategy when allowed access to post-quantum ``states''. The red point on the surface of the Bloch sphere represents the optimal message state $\ket{\psi_{0,0}}$, achieving of a value of $0.79$ for the Torpedo game. The black point representing $\frac{1}{2}\left(\mathbb{I}-(X+Y+Z)\right)$ is not a valid density matrix, but achieves a value of 1 in the torpedo game. }
        \label{fig:Postquantum}
\end{figure}

\sloppy The average probability of success for the Casaccino et al.~QRAC, see \refequ{GalvaoQRAC}, can be maximized by using a post-quantum ``state'' of the form $\Pi_1^{k_1}+\Pi_2^{k_2}+\ldots+ \Pi_{d+1}^{k_{d+1}}-\mathbb{I}$, where scare quotes reflect the fact that, although it is Hermitian and has unit trace, its spectrum is not necessarily nonnegative. In fact the ``state'' above is a phase point operator, $A_{\pmb{k}}$, for one of Wootters' discrete Wigner functions. Since phase point operators obey $\Tr(A\mathbb{I})=1$ and $\Tr(AA)=d$ then $\frac{1}{(d+1)d^{d+1}}\sum_{\pmb{k}\in\mathbb{Z}_d^{d+1}}\Tr\left[A_{\pmb{k}}(A_{\pmb{k}}+\mathbb{I})\right] =1$. In other words, there is a perfect strategy by using post-quantum states. Seen in this way, phase point operators in a $(d+1,1)_d$ QRAC scenario are similar to Popescu-Rohrlich \cite{popescu1994quantum} boxes in the CHSH scenario. As seen above, our Torpedo Game has a perfect strategy within quantum mechanics for all odd power-of-prime dimensions, by construction. In contrast, we saw that the qubit Torpedo Game only has quantum value of
roughly $0.79$.
To reach a perfect strategy, we must once again use a phase point operator as the non-physical ``state'' that Alice sends to Bob, see Fig.~\ref{fig:Postquantum}.

\subsection{Optimal Classical Strategies}
\label{sec:classical_strat}

In what follows, we describe an encoding map $\Ec = \{p_{\mathcal{E}}(\cdot|x,z)\}_{x,z}$ as specifying a probability distribution over messages $j \in \mathbb{Z}_d$  for each combination of inputs $x,z \in \mathbb{Z}_d$.
Similarly a decoding map $\Dc = \{ p_{\mathcal{D}}(\cdot|j,q) \}_{j,q}$ specifies a probability distribution over outputs $c \in \mathbb{Z}_d$ for each combination of a message and question, $j,q \in \mathbb{Z}_d$ respectively.

Combining an encoding
$\Ec$
and a decoding
$\Dc$
results in an empirical behaviour $e = \{ p_{e}(\cdot|x,z,q) \}_{x,z,q}$.
This is a set of probability distributions over outputs $c \in \mathbb{Z}_d$, one for each combination of the referee variables $x,z,q \in \mathbb{Z}_d$,
such that
\begin{equation}
p_{e}(c |x,z,q) = \sum_{j \in \Z_d} \, p_{\mathcal{D}}( c |j,q) \, p_{\mathcal{E}}(j|x,z) \, .
    \label{eq:cl_proba_emp}
\end{equation}
By comparison, quantum mechanical empirical behaviours arise via the Born rule: $p_e(c|x,z,q) = \Tr(\rho_{x,z}\Pi_q^c)$.

Assuming the referee variables to be uniformly distributed, a strategy has a winning probability given in terms of its empirical probabilities as
\[
\frac{1}{d^2(d+1)} \sum_{x,z,q} p_e (w_q(x,z) \mid x,z,q) \, .
\]
The classical value of the Torpedo Game for dimension $d$ can thus be expressed as
\begin{equation}
    \theta^C_{d} = \Max{\Ec,\Dc}
    \bigg[ \frac{1}{d^2(d+1)} \sum_{x,z,q}
    \, p_e ( w_q(x,z) \mid x,z,q) \bigg] \, .
    \label{eq:cval}
\end{equation}

For evaluation of this expression note that it suffices to consider deterministic encodings and decodings.
In the presence of shared randomness,
nondeterministic strategies can always be obtained as convex combinations of deterministic ones and the expression is convex linear \cite{gallego2010device}.
Furthermore, for each encoding there exists a decoding that is optimal with respect to it.
This fact was also observed for one-way communication tasks with messages of bounded dimension in \cite{saha2019state}.
Thus it is possible to evaluate the classical value by enumerating over deterministic encodings only.

\begin{proposition}
The classical value of an information retrieval task can be expressed as a maximum over encodings as
\begin{equation}
	\theta^c=
	\max_{\Ec}
	\bigg[
	\frac{1}{d^2(d+1)}\sum_{j,q}
	\max_c \sum_{\substack{ (x,z) \text{ s.t.} \\ c \in w_q(x,z)}} p_{\mathcal{E}}(j|x,z)
	 \bigg] .
\label{eq:classical_formula}
\end{equation}
\end{proposition}

\begin{proof}
Starting from \refequ{cval},
\begin{align*}
    \theta^c &= \Max{\Ec,\Dc}
    \bigg[ \frac{1}{d^2(d+1)} \, \sum_{x,z,q}
    \, p_e ( w_q(x,z) \mid x,z,q) \bigg] \\
    &= \Max{\Ec,\Dc}
    \bigg[ \frac{1}{d^2(d+1)} \, \sum_{x,z,q} \, \sum_{c \in w_q(x,z)}
    \, p_e ( c \mid x,z,q) \bigg] \\
    &= \Max{\Ec,\Dc}
    \bigg[ \frac{1}{d^2(d+1)} \, \sum_{q,c} \, \sum_{\substack{(x,z) \text{s.t. } \\ c \in w_q(x,z)}}
    \, p_e ( c \mid x,z,q) \bigg] \\
    &= \Max{\Ec,\Dc}
    \bigg[ \frac{1}{d^2(d+1)} \, \sum_{j,q,c} \, \sum_{\substack{(x,z) \text{s.t. } \\ c \in w_q(x,z)}}
    \, p_\Dc (c | j,q) \, p_\Ec ( j \mid x,z) \bigg] \\
    &= \Max{\Ec} \bigg[ \frac{1}{d^2(d+1)} \, \sum_{j,q} \, \max_c \sum_{\substack{(x,z) \text{s.t. } \\ c \in w_q(x,z)}}
    \, p_\Ec ( j \mid x,z) \bigg] \, ,
\end{align*}
where the last line follows by using a deterministic decoding that is optimal with respect to the encoding.
\end{proof}

A useful way of representing any deterministic encoding is as a colouring of the $d \times d$ affine plane using no more than $d$ colours.
Observe that a deterministic encoding can alternatively be expressed as a function $f_\Ec : \mathbb{Z}_d \times \mathbb{Z}_d \rightarrow \mathbb{Z}_d$, where $f_\Ec(x,z)$ is the message dit to be sent (with probability $1$) given inputs $x,z$.
Thinking of the inputs as coordinates in the $d \times d$ affine plane
a deterministic encoding is equivalent to a partition of the plane into no more than $d$ equivalence classes,
or a colouring using no more than $d$ colours.

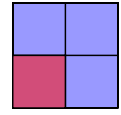
\begin{figure}[htbp]
        \centering
        \scalebox{\myscale}{    \centering
    \begin{tikzpicture}[scale=1]

        \draw [thick, fill=blue!40] (0,0) grid (2,2) rectangle (0,0);
        \draw [thick, fill=purple!70] (0,0) rectangle (1,1);

    \end{tikzpicture}}
        \caption{
        An optimal classical strategy for the $d=2$ Torpedo Game. Alice uses her bit of communication to indicate in which class of the partition that she finds herself. Classes are represented here by colours.
        }
        \label{fig:cl_d2}
    \end{figure}

\paragraph{Optimal Strategies for $d=2$ and $d=3$.}
In general there are $d^{d^2}$ partitions of a $d \times d$ grid.
For low dimensions the expression in \refequ{classical_formula} can be evaluated by exhaustive search over partitions.
For dimension $2$ and $3$ we find
\begin{equation}
\label{eq:bounds}
    \theta^{C}_{d=2} = \frac{3}{4} \quad \text{ and } \quad \theta^{C}_{d=3} = \frac{11}{12} \, .
\end{equation}
Example of strategies that attains these values are depicted below in \refig{cl_d2} and in \refig{cl_d3}.

    \begin{figure}[htpb]
        \centering

	
	\begin{tikzpicture}[scale=.6]
	
	\draw [thick, fill=blue!40] (0,0) grid (3,3) rectangle (0,0);
	
	\draw [thick, fill=purple!70] (0,0) rectangle (1,1);
	\draw [thick, fill=purple!70] (1,0) rectangle (2,1);
	\draw [thick, fill=purple!70] (2,1) rectangle (3,2);
	\draw [thick, fill=purple!70] (2,1) rectangle (3,2);

	\draw [thick, fill=green!30] (0,2) rectangle (1,3);
	\draw [thick, fill=green!30] (1,2) rectangle (2,3);
	\draw [thick, fill=green!30] (1,1) rectangle (2,2);
	
	\end{tikzpicture}
	
\bigskip
\begin{tikzpicture}[scale=.15]


\draw [fill=gray!20] (-1,-4) rectangle (2,-1);
\draw [fill=purple!70] (-1,-2) rectangle (2,-1);
\draw [thick, fill=gray!20] (-1,-4) grid (2,-1);

\draw [fill=gray!20] (7,-4) rectangle (10,-1);
\draw [fill=green!30] (7,-4) rectangle (10,-3);
\draw [thick, fill=gray!20] (7,-4) grid (10,-1);

\draw [fill=gray!20] (3,-4) rectangle (6,-1);
\draw [fill=black!70] (3,-3) rectangle (6,-2);
\draw [thick, fill=gray!20] (3,-4) grid (6,-1);


\draw [fill=gray!20] (16,-4) rectangle (19,-1);
\draw [fill=blue!40] (17,-4) rectangle (18,-1);
\draw [thick, fill=gray!20] (16,-4) grid (19,-1);

\draw [fill=gray!20] (12,-4) rectangle (15,-1);
\draw [fill=black!70] (12,-4) rectangle (13,-1);
\draw [thick, fill=gray!20] (12,-4) grid (15,-1);

\draw [fill=gray!20] (20,-4) rectangle (23,-1);
\draw [fill=green!30] (22,-4) rectangle (23,-1);
\draw [thick, fill=gray!20] (20,-4) grid (23,-1);


\draw [fill=gray!20] (29,-4) rectangle (32,-1);
\draw [fill=purple!70] (29,-2) rectangle (30,-1);
\draw [fill=purple!70] (30,-3) rectangle (31,-2);
\draw [fill=purple!70] (31,-3) rectangle (32,-4);
\draw [thick, fill=gray!20] (29,-4) grid (32,-1);

\draw [fill=gray!20] (25,-4) rectangle (28,-1);
\draw [fill=blue!40] (25,-4) rectangle (26,-3);
\draw [fill=blue!40] (26,-2) rectangle (27,-1);
\draw [fill=blue!40] (27,-3) rectangle (28,-2);
\draw [thick, fill=gray!20] (25,-4) grid (28,-1);

\draw [fill=gray!20] (33,-4) rectangle (36,-1);
\draw [fill=green!30] (33,-3) rectangle (34,-2);
\draw [fill=green!30] (34,-4) rectangle (35,-3);
\draw [fill=green!30] (35,-2) rectangle (36,-1);
\draw [thick, fill=gray!20] (33,-4) grid (36,-1);


\draw [fill=gray!20] (46,-4) rectangle (49,-1);
\draw [fill=black!70] (46,-4) rectangle (47,-3);
\draw [fill=black!70] (47,-3) rectangle (48,-2);
\draw [fill=black!70] (48,-2) rectangle (49,-1);
\draw [thick, fill=gray!20] (46,-4) grid (49,-1);

\draw [fill=gray!20] (38,-4) rectangle (41,-1);
\draw [fill=purple!70] (38,-3) rectangle (39,-2);
\draw [fill=purple!70] (39,-2) rectangle (40,-1);
\draw [fill=purple!70] (40,-4) rectangle (41,-3);
\draw [thick, fill=gray!20] (38,-4) grid (41,-1);

\draw [fill=gray!20] (42,-4) rectangle (45,-1);
\draw [fill=blue!40] (42,-2) rectangle (43,-1);
\draw [fill=blue!40] (43,-4) rectangle (44,-3);
\draw [fill=blue!40] (44,-3) rectangle (45,-2);
\draw [thick, fill=gray!20] (42,-4) grid (45,-1);

\end{tikzpicture}
        \caption{
        An optimal classical strategy for the $d=3$ Torpedo Game.
        Alice uses her dit of communication to indicate in which equivalence class (represented by same coloured cells) of the large grid partition she finds herself.
        The smaller grids (cf.\ Fig.~\ref{fig:d3_torpedo_firing}) show where Bob chooses to shoot, given a direction and a colour.
        For the first direction, when asked to shoot horizontally in the grid, notice that Bob may avoid Alice with certainty if she is in either of the red or green partitions.
        Lines that avoid Alice with certainty are depicted in the corresponding colour,
        whereas black lines intersect with Alice's position with probability $\frac{1}{3}$.
        Overall, this strategy wins the Torpedo Game with probability
$\frac{1}{4}(\frac{8}{9}+\frac{8}{9}+1+\frac{8}{9}) = \frac{11}{12}$.
        }
        \label{fig:cl_d3}
    \end{figure}

\begin{figure}[htbp]
        \centering
\begin{tikzpicture}[scale=.45]

\draw [thick, fill=blue!40] (0,0) grid (5,5) rectangle (0,0);

\draw [thick, fill=brown!60] (0,0) rectangle (1,1);
\draw [thick, fill=brown!60] (1,0) rectangle (2,1);
\draw [thick, fill=brown!60] (0,2) rectangle (1,3);
\draw [thick, fill=brown!60] (0,4) rectangle (1,5);
\draw [thick, fill=brown!60] (4,1) rectangle (5,2);

\draw [thick, fill=green!30] (0,1) rectangle (1,2);
\draw [thick, fill=green!30] (2,2) rectangle (3,3);
\draw [thick, fill=green!30] (2,3) rectangle (3,4);
\draw [thick, fill=green!30] (3,0) rectangle (4,1);
\draw [thick, fill=green!30] (3,1) rectangle (4,2);

\draw [thick, fill=cyan!70] (2,0) rectangle (3,1);
\draw [thick, fill=cyan!70] (1,2) rectangle (2,3);
\draw [thick, fill=cyan!70] (1,4) rectangle (2,5);
\draw [thick, fill=cyan!70] (3,4) rectangle (4,5);
\draw [thick, fill=cyan!70] (4,3) rectangle (5,4);

\draw [thick, fill=purple!70] (1,1) rectangle (2,2);
\draw [thick, fill=purple!70] (3,2) rectangle (4,3);
\draw [thick, fill=purple!70] (3,3) rectangle (4,4);
\draw [thick, fill=purple!70] (2,4) rectangle (3,5);
\draw [thick, fill=purple!70] (4,4) rectangle (5,5);

\end{tikzpicture}
\bigskip

\begin{tikzpicture}[scale=.18]

\draw[fill = green!30] (0,4) -- (0,5) -- (1,4);
\draw[fill = green!30] (1,4) -- (1,5) -- (2,4);
\draw[fill = green!30] (2,4) -- (2,5) -- (3,4);
\draw[fill = green!30] (3,4) -- (3,5) -- (4,4);
\draw[fill = green!30] (4,4) -- (4,5) -- (5,4);
\draw[fill = blue!40] (0,5) -- (1,5) -- (1,4);
\draw[fill = blue!40] (1,5) -- (2,5) -- (2,4);
\draw[fill = blue!40] (2,5) -- (3,5) -- (3,4);
\draw[fill = blue!40] (3,5) -- (4,5) -- (4,4);
\draw[fill = blue!40] (4,5) -- (5,5) -- (5,4);
\draw[fill= purple!70] (0,0) rectangle (5,1);
\draw[fill= cyan!70] (0,1) rectangle (5,2);
\draw[fill= brown!60] (0,3) rectangle (5,4);
\node[below] (notx) at (2.5,0) {\scriptsize{$\neg (x)$}};
\draw (0,0) grid (5,5);

\draw[fill = purple!70] (7,0) -- (8,0) -- (7,1);
\draw[fill = purple!70] (7,1) -- (8,1) -- (7,2);
\draw[fill = purple!70] (7,2) -- (8,2) -- (7,3);
\draw[fill = purple!70] (7,3) -- (8,3) -- (7,4);
\draw[fill = purple!70] (7,4) -- (8,4) -- (7,5);
\draw[fill = cyan!70] (7,1) -- (8,1) -- (8,0);
\draw[fill = cyan!70] (7,2) -- (8,2) -- (8,1);
\draw[fill = cyan!70] (7,3) -- (8,3) -- (8,2);
\draw[fill = cyan!70] (7,4) -- (8,4) -- (8,3);
\draw[fill = cyan!70] (7,5) -- (8,5) -- (8,4);
\draw[fill = green!30] (8,0) rectangle (9,5);
\draw[fill = blue!40] (10,0) rectangle (11,5);
\draw[fill = brown!60] (9,0) rectangle (10,5);
\draw (7,0) grid (12,5);
\node[below] (notz) at (9.5,0) {\scriptsize{$\neg (-z)$}};

\draw[fill = blue!40] (14,0) rectangle (15,1);
\draw[fill = blue!40] (15,4) rectangle (16,5);
\draw[fill = blue!40] (16,3) rectangle (17,4);
\draw[fill = blue!40] (17,2) rectangle (18,3);
\draw[fill = blue!40] (18,1) rectangle (19,2);
\draw[fill = cyan!70] (14,1) rectangle (15,2);
\draw[fill = cyan!70] (15,0) rectangle (16,1);
\draw[fill = cyan!70] (16,4) rectangle (17,5);
\draw[fill = cyan!70] (17,3) rectangle (18,4);
\draw[fill = cyan!70] (18,2) rectangle (19,3);
\draw[fill = green!30] (14,2) rectangle (15,3);
\draw[fill = green!30] (15,1) rectangle (16,2);
\draw[fill = green!30] (16,0) rectangle (17,1);
\draw[fill = green!30] (17,4) rectangle (18,5);
\draw[fill = green!30] (18,3) rectangle (19,4);
\draw[fill = brown!60] (14,3) rectangle (15,4);
\draw[fill = brown!60] (15,2) rectangle (16,3);
\draw[fill = brown!60] (16,1) rectangle (17,2);
\draw[fill = brown!60] (17,0) rectangle (18,1);
\draw[fill = brown!60] (18,4) rectangle (19,5);
\draw[fill = purple!70] (14,4) rectangle (15,5);
\draw[fill = purple!70] (15,3) rectangle (16,4);
\draw[fill = purple!70] (16,2) rectangle (17,3);
\draw[fill = purple!70] (17,1) rectangle (18,2);
\draw[fill = purple!70] (18,0) rectangle (19,1);
\draw (14,0) grid (19,5);
\node[below] (notxz) at (16.5,0) {\scriptsize{$\neg (x-z)$}};

\draw[fill = brown!60] (21,3) rectangle (22,4);
\draw[fill = brown!60] (22,1) rectangle (23,2);
\draw[fill = brown!60] (23,4) rectangle (24,5);
\draw[fill = brown!60] (24,2) rectangle (25,3);
\draw[fill = brown!60] (25,0) rectangle (26,1);
\draw[fill = green!30] (21,4) rectangle (22,5);
\draw[fill = green!30] (22,2) rectangle (23,3);
\draw[fill = green!30] (23,0) rectangle (24,1);
\draw[fill = green!30] (24,3) rectangle (25,4);
\draw[fill = green!30] (25,1) rectangle (26,2);
\draw[fill = purple!70] (21,0) rectangle (22,1);
\draw[fill = purple!70] (22,3) rectangle (23,4);
\draw[fill = purple!70] (23,1) rectangle (24,2);
\draw[fill = purple!70] (24,4) rectangle (25,5);
\draw[fill = purple!70] (25,2) rectangle (26,3);
\draw[fill = blue!40] (21,1) rectangle (22,2);
\draw[fill = blue!40] (22,4) rectangle (23,5);
\draw[fill = blue!40] (23,2) rectangle (24,3);
\draw[fill = blue!40] (24,0) rectangle (25,1);
\draw[fill = blue!40] (25,3) rectangle (26,4);
\draw[fill = cyan!70] (21,2) rectangle (22,3);
\draw[fill = cyan!70] (22,0) rectangle (23,1);
\draw[fill = cyan!70] (23,3) rectangle (24,4);
\draw[fill = cyan!70] (24,1) rectangle (25,2);
\draw[fill = cyan!70] (25,4) rectangle (26,5);
\draw (21,0) grid (26,5);
\node[below] (notxz2) at (23.5,0) {\scriptsize{$\neg (2x-z)$}};

\draw[fill = green!30] (28,2) rectangle (29,3);
\draw[fill = green!30] (29,4) rectangle (30,5);
\draw[fill = green!30] (30,1) rectangle (31,2);
\draw[fill = green!30] (31,3) rectangle (32,4);
\draw[fill = green!30] (32,0) rectangle (33,1);
\draw[fill = brown!60] (28,1) rectangle (29,2);
\draw[fill = brown!60] (29,3) rectangle (30,4);
\draw[fill = brown!60] (30,0) rectangle (31,1);
\draw[fill = brown!60] (31,2) rectangle (32,3);
\draw[fill = brown!60] (32,4) rectangle (33,5);
\draw[fill = blue!40] (28,0) rectangle (29,1);
\draw[fill = blue!40] (29,2) rectangle (30,3);
\draw[fill = blue!40] (30,4) rectangle (31,5);
\draw[fill = blue!40] (31,1) rectangle (32,2);
\draw[fill = blue!40] (32,3) rectangle (33,4);
\draw[fill = purple!70] (28,3) rectangle (29,4);
\draw[fill = purple!70] (29,0) rectangle (30,1);
\draw[fill = purple!70] (30,2) rectangle (31,3);
\draw[fill = purple!70] (31,4) rectangle (32,5);
\draw[fill = purple!70] (32,1) rectangle (33,2);
\draw[fill = cyan!70] (28,4) rectangle (29,5);
\draw[fill = cyan!70] (29,1) rectangle (30,2);
\draw[fill = cyan!70] (30,3) rectangle (31,4);
\draw[fill = cyan!70] (31,0) rectangle (32,1);
\draw[fill = cyan!70] (32,2) rectangle (33,3);
\draw (28,0) grid (33,5);
\node[below] (notxz3) at (30.5,0) {\scriptsize{$\neg (3x-z)$}};

\draw[fill = cyan!70] (35,2) rectangle (36,3);
\draw[fill = cyan!70] (36,3) rectangle (37,4);
\draw[fill = cyan!70] (37,4) rectangle (38,5);
\draw[fill = cyan!70] (38,0) rectangle (39,1);
\draw[fill = cyan!70] (39,1) rectangle (40,2);
\draw[fill = blue!40] (35,0) rectangle (36,1);
\draw[fill = blue!40] (36,1) rectangle (37,2);
\draw[fill = blue!40] (37,2) rectangle (38,3);
\draw[fill = blue!40] (38,3) rectangle (39,4);
\draw[fill = blue!40] (39,4) rectangle (40,5);
\draw[fill = purple!70] (35,1) rectangle (36,2);
\draw[fill = purple!70] (36,2) rectangle (37,3);
\draw[fill = purple!70] (37,3) rectangle (38,4);
\draw[fill = purple!70] (38,4) rectangle (39,5);
\draw[fill = purple!70] (39,0) rectangle (40,1);
\draw[fill = brown!60] (35,3) rectangle (36,4);
\draw[fill = brown!60] (36,4) rectangle (37,5);
\draw[fill = brown!60] (37,0) rectangle (38,1);
\draw[fill = brown!60] (38,1) rectangle (39,2);
\draw[fill = brown!60] (39,2) rectangle (40,3);
\draw[fill = green!30] (35,4) rectangle (36,5);
\draw[fill = green!30] (36,0) rectangle (37,1);
\draw[fill = green!30] (37,1) rectangle (38,2);
\draw[fill = green!30] (38,2) rectangle (39,3);
\draw[fill = green!30] (39,3) rectangle (40,4);
\draw (35,0) grid (40,5);
\node[below] (notxz4) at (37.5,0) {\scriptsize{$\neg (4x-z)$}};

\end{tikzpicture}
        \caption{
        A perfect classical strategy for the $d=5$ Torpedo Game. 
        As \refig{cl_d3}, same coloured cells belong to the same partition.
        The lines that avoid Alice are depicted below for every questions Bob can be asked.
        }
        \label{fig:cl_d5}
    \end{figure}
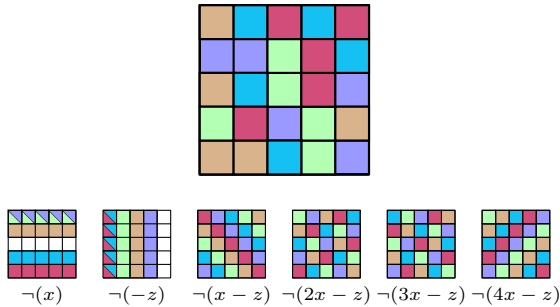

\paragraph{Optimal Strategies for $d=5$ and beyond.}
As $d$ increases it quickly becomes infeasible to perform an exhaustive search over all partitions.
We have, however, found perfect classical strategies, i.e.\ strategies that win with probability $1$, for $d=5$ (see \refig{cl_d5}) up to $d=23$.
This leads us to conjecture that there exists a perfect classical strategy for all $d>5$,
\begin{equation}
    \textbf{Conjecture:}\quad \theta^{C}_{d \geq 5} = 1. \label{eq:PerfectClassical}
\end{equation}

\subsection{Comparison of Quantum and Classical Game Values}
Recall the optimal quantum values established in \refsec{qstrat},
\begin{equation}
    \theta^{Q}_ {d = 2} \simeq 0.79 \quad \text{ and } \quad \theta^{Q}_{d \geq 3}= 1 \, . \label{eq:QuantumValues}
\end{equation}
Comparing these with the classical bounds from \refsec{classical_strat} we obtain the ratios
\begin{equation}
    \frac{\theta^{Q}_ {d = 2}}{\theta^{C}_ {d = 2}} \simeq 1.053 \quad \text{ and } \quad \frac{\theta^{Q}_ {d = 3}}{\theta^{C}_ {d = 3}} \simeq 1.091 \, .
\end{equation}
By comparison, it was shown in \cite{tavakoli2015quantum} that the classical and quantum values of the $(4,1)_3$ (Q)RAC are $\frac{16}{27}$ and $0.637$, respectively, giving a ratio of $\theta^{Q}_ {d = 3} / \theta^{C}_ {d = 3} \simeq 1.075$.
Accordingly, we note that the $d=3$ Torpedo Game admits a greater quantum-over-classical advantage than the standard random access coding task whose optimal QRAC also exploits the $4$ mutually unbiased bases available in dimension $3$.

    \section{Characterising and Quantifying Contextuality in Information Retrieval Tasks}
        \label{sec:context}

\label{sec:contextuality}

When quantum advantage is observed in a bounded-memory information retrieval task like a (Q)RAC task or the Torpedo Game, it highlights a difference between the information carrying capacities of qudits compared to dits for a fixed dimension.
It can be remarked that such a difference is a consequence of the different geometries of the respective state spaces.
In this section, however, we seek a sharper, quantified analysis of the source of the advantage in terms of contextuality.
We follow the sequential approach to contextuality of \cite{mansfield2018quantum}, as a behavioural characteristic that may arise in sequences of operations and is distinct from that of Spekkens \cite{spekkens2005contextuality}.
For the bounded-memory regime we are interested in sequential contextuality also matches the characteristic introduced by \.Zukowski in \cite{zukowski2014temporal}.
While \cite{mansfield2018quantum} was concerned specifically with sequences of transformations,
we take a broader view that applies also to prepare-and-measure scenarios.

The study of contextuality arose in quantum foundations,
where a major theme is the attempt to understand empirical behaviours that may appear non-intuitive from a classical perspective, e.g.\ the EPR paradox \cite{einstein1935can}.
The typical approach is to look for a description of physical systems at a deeper level than the quantum one at which more classically intuitive properties may be restored.
Such a description is usually formalised as a hidden variable model for the behaviour (sometimes also referred to as an ontological model \cite{spekkens2005contextuality}).
The great significance of the celebrated no-go theorems of quantum foundations,
like Bell's Theorem \cite{bell1964einstein} and the Bell--Kochen--Specker Theorem \cite{bell1966,ks},
was to prove that certain `non-classical' features of the empirical behaviours of quantum systems are necessarily inherited by any underlying model.

Non-classical features of quantum systems like contextuality are also increasingly investigated for their practical utility.
For instance, in previous work involving present authors, contextuality of the Bell--Kochen--Specker kind was shown to be a prerequisite for quantum speed-up \cite{howard2014contextuality} and to quantify quantum-over-classical advantage in a variety of informational tasks \cite{abramsky2017contextual}.

Bell--Kochen--Specker contextuality essentially concerns the statistics that arise under varied measurements on a physical system.
In contrast, our notion of contextuality concerns the statistics that arise from sequences of operations -- preparations, transformations and measurements -- all of which can vary.
As a behavioural feature sequential contextuality signifies the absence of a hidden variable model that respects the sequential structure of these operations.

Rather than focusing on characteristics that must be inherited by all hidden variable models,
as is common in foundational works,
we also take a practical perspective and shift focus to characteristics that must be inherited by bounded-memory models --
a constraint that matches the informational problem at hand.
In this respect, the significance of sequential contextuality in what follows can be viewed through a practical rather than a foundational lens, as a characteristic that quantifies quantum advantage.

\subsection{Empirical Behaviours and Hidden Variable Models}
\label{sec:emp_behaviours}

Recall from Section \ref{sec:classical_strat} that any strategy for an information retrieval task gives rise to an empirical behaviour $e = \{ p_e (\cdot | i,q) \}_{i,q}$.
In other words, for each combination of input string $i \in \mathbb{Z}_d^n$ and question $q \in Q$
there is a resulting probability distribution over outputs.
This is true regardless of whether the strategy is classical, quantum, post-quantum, or other.
The combination of input string and question will be referred to as the context.
This description of empirical behaviour echoes the formalism of empirical models for measurement scenarios in \cite{ab}.

Given an empirical behaviour, one can ask whether it can be simulated by a bounded-memory hidden variable model.
In particular we will be interested in models that respect the sequential structure of the strategy,
bearing in mind that the inputs and questions specify a sequence of operations: either preparations, transformations, or measurements.
Our main focus is on prepare-and-measure scenarios, in which sequences arise from a combination of a preparation and a measurement.
To match the constraints of information retrieval tasks, memory is bounded by the dimension of the message string.

In a bounded-memory model, the hidden variable is restricted to take values in $\mathbb{Z}_d$ with $d$ fixed by the communication scenario.
A state preparation $P$ is modelled by a probability distribution $p_\pi(\cdot \mid P)$ over the hidden variable space $\mathbb{Z}_d$.
Similarly, a {measurement} $M$ is modelled by a family of probability distributions $\{ p_\mu(\cdot \mid \lambda, M) \}_\lambda$ over the outcome space, which to match the communication scenario is also $\mathbb{Z}_d$.
A hidden variable model, e.g.\ for a prepare-and-measure sequence of operations $M_q \circ P_{x,z}$ as in \refig{PAM_Transform}, simulates an empirical behaviour as

\begin{equation}
p_e(\cdot \mid x,z,q ) = \\ \sum_{\lambda} \, p_\mu(\cdot \mid \lambda, M_q) \, p_\pi(\lambda \mid P_{x,z}) \, .
    \label{eq:pamprealise}
\end{equation}

With this in mind, the bounded-memory classical strategies of \refsec{optima} can also be interpreted as bounded-memory hidden variable models themselves.
To see this, note that in \refequ{cl_proba_emp} encoding corresponds to hidden-variable preparation and decoding to hidden-variable measurement.

While the above description makes contact with the strategies and empirical behaviours of \refsec{optima}, it will be convenient for the remainder of this section to use a simplified notation.
Let us work in the real vector space with basis given by the hidden variable states $\Z_d$.
For all preparations $P_{x,z}$, the probability distribution $p_\pi(\cdot \mid P_{x,z})$ will be more concisely denoted as a probability vector $\pmb{\lambda}_{x,z}$.
For measurements, we express $\{ p_\mu(\cdot \mid \lambda, M) \}_\lambda$ more concisely as a set $\left\{ \pmb{v}_q^c \right\}_{c \in \Z^d}$ of positive real vectors, one for each outcome $c\in \Z_d$.
These satisfy the property that for all $q$, $\sum_{c \in \Z_d} \pmb{v}_q^c=\pmb{1}$, the vector all of whose entries are $1$.
If we also denote by $e_{x,z,q} := p_e(\cdot \mid x,z,q)$ the empirical probability vector over outcomes $\mathbb{Z}_d$, then \refequ{pamprealise} can be rewritten for each outcome $c \in \Z_d$ in simplified notation as the dot product
\begin{align}
    e_{x,z,q}(c) &= \pmb{v}_q^c \cdot \bm{\lambda}_{x,z} \, .
    \label{eq:pamrealise}
\end{align}

As our focus is on prepare-and-measure scenarios, we have not discussed hidden-variable modelling
of transformations. This is discussed in the appendix, \refsec{transformation}.

\subsection{Sequential Contextuality in Information Retrieval Tasks}

An empirical behaviour is sequential noncontextual if it admits a hidden variable model that:
(i) preserves a modular sequential description of operations, and
(ii) the hidden variable representation of operations is context-independent.

These assumptions have been implicitly built into the above definition of hidden-variable models.
For (i), note that each operation has an individual description at the hidden-variable level.
For example, to obtain predictions for a prepare-and-measure experiment we compose the individual hidden-variable descriptions of the preparation and of the measurement, as in \refequ{pamrealise}.
And for (ii), note for example that regardless of which context the preparation $P_{x,z}$ appears in it should be modelled by the same vector $\bm{\lambda}_{x,z}$.
One could relax these assumptions, in which case it would become trivial to find a hidden-variable model for any behaviour,
but it would also entail giving up the intuitive sense of what the model means.

If an empirical behaviour does not admit a sequential noncontextual hidden-variable model it is said to be sequential contextual.
In this article we will only be considering sequential contextuality with respect to bounded-memory models, though the definition may be applied more generally.

A useful intuition for sequential contextuality is that, within the memory constraints, for any faithful model of the behaviour the whole (the description of the context) is more than the composition of its parts (the description of the individual operations).
A contextual model would always involve additional memory and communication to track the context, which would be outside of the constraints of the task --
involving, e.g., a contextuality demon analogous to Maxwell's demon in thermodynamics.
Indeed it was shown in \cite{henaut2018tsirelson} that a related characteristic incurs a simulation cost as measured by Landauer erasure.

\subsection{Quantifying Contextuality via the Contextual Fraction}
For any fixed $(n,m)_d$ communication scenario empirical behaviours are closed under context-wise convex combinations --
a property that is inherited from probability distributions.
In operational terms, if shared randomness is used to choose between several strategies the result is still an empirical behaviour.

For any empirical behaviour $e$, we can consider convex decompositions of the form
\begin{equation}\label{eq:edecomp}
e = \omega e^{\mathrm{NC}} + (1 - \omega) e' \, ,
\end{equation}
where $e^{\mathrm{NC}}$ and $e'$ are empirical behaviours for the same task and $e^{\mathrm{NC}}$ is noncontextual.
The maximum of $\omega$ over all such decompositions is referred to as the noncontextual fraction of $e$, written $\mathrm{NCF}(e)$.
Similarly, the {contextual fraction} of $e$ is $\mathrm{CF}(e) := 1 - \mathrm{NCF}(e)$.

This provides a measure of contextuality in the interval $[0,1]$, where $\mathrm{CF}(e)=0$ indicates that $e$ is noncontextual, $\mathrm{CF}(e)>0$ indicates that $e$ is contextual, and $\mathrm{CF}(e)=1$ indicates that $e$ is maximally contextual (also referred to as strong contextuality).

The contextual fraction was used as a measure for sequential contextuality in \cite{mansfield2018quantum}.
It extends a natural measure for Bell--Kochen--Specker contextuality \cite{ab}, which itself generalises measures based on Bell-inequality violations \cite{abramsky2017contextual}.

\subsection{Quantified Contextual Advantage in Information Retrieval Tasks}

\begin{figure}[h!]
    \centering
    \scalebox{\myscale}{    \usetikzlibrary{arrows,3d,fadings,shapes,calc,decorations.pathreplacing,decorations.markings,positioning}
\def\tkzscl{0.8}

\begin{tikzpicture}[scale=\tkzscl]

\node (up) at (0,1) {};
\node (down) at (0,-1) {};
\node (left) at (-1,0) {};
\node (right) at (1,0) {};

\node (a1) at (0,0) {};
\node (a2) at (6,2) {};
\node [inner sep=.0pt] (a3) at (6,1) {};

\node (b1) at (8,0) {};
\node (b2) at (12,2) {};
\node [inner sep=.0pt] (b3) at (8,1) {};

\draw[rounded corners,dotted,fill=gray!20] (a1) rectangle (a2);
\node (am) at (2,1) {};
\node (E) at ($(am)+(right)$) {\huge $\mathcal{E}$};
\node [inner sep=0 pt] (d1) at ($(am)+1.5*(up)+(right)$) {};
\node [inner sep=0 pt] (d2) at ($(am)+1.5*(up)+3*(right)$) {};
\draw [double distance=1.5pt] (d1) to ($(d1)+.5*(down)$);
\draw [double distance=1.5pt] (d2) to ($(d2)+.5*(down)$);
\node (d1texte) at ($(d1)+.3*(up)$) {\large $x$};
\node (d2texte) at ($(d2)+.3*(up)$) {\large $z$};

\draw[rounded corners,dotted,fill=gray!20,dotted] (b1) rectangle (b2);
\node (bm) at (10,1) {\huge $\Dc$};
\node [inner sep=0 pt] (j) at ($(bm)+1.5*(up)+(left)$) {};
\draw [double distance=1.5pt] (j) to ($(j)+.5*(down)$);
\node (jtexte) at ($(j)+.3*(up)+.5*(right)$) {$q \in \{\infty,0,1,2, \dots, d-1 \}$};
\node [inner sep=0 pt] (o) at ($(bm)+2*(right)$) {};
\draw [double distance=1.5pt] (o) to ($(o)+(right)$);
\node (otexte) at ($(o)+1.3*(right)$) {$c$};

\draw[ultra thick] (a3) to (b3); 
\draw[densely dotted, red] ($.5*(a2)+.5*(b1)+3*(up)$) to ($.5*(a2)+.5*(b1)+2*(down)$);
\node (Alice) at ($.5*(a3)+.5*(b3)+2.5*(up)+.8*(left)$) {\textcolor{red}{Alice}};
\node (Bob) at ($.5*(a3)+.5*(b3)+2.5*(up)+.8*(right)$) {\textcolor{red}{Bob}};
\node (psi) at ($.5*(a2)+.5*(b1)+.5*(up)$) {\large $j$};

\end{tikzpicture}}
    \scalebox{\myscale}{    \usetikzlibrary{arrows,3d,fadings,shapes,calc,decorations.pathreplacing,decorations.markings,positioning}
\def\tkzscl{0.8}

\begin{tikzpicture}[scale=\tkzscl]

\node (up) at (0,1) {};
\node (down) at (0,-1) {};
\node (left) at (-1,0) {};
\node (right) at (1,0) {};

\node (a1) at (0,0) {};
\node (a2) at (6,2.5) {};
\node [inner sep=.0pt] (a3) at (6,1) {};

\node (b1) at (8,0) {};
\node (b2) at (12,2.5) {};
\node [inner sep=.0pt] (b3) at (8,1) {};

\draw[rounded corners,dotted,fill=gray!20] (a1) rectangle (a2);
\node (am) at (2,1) {};
\node (E) at ($(am)+.2*(left)+(up)$) {\huge $\mathcal{E}$};
\node [inner sep=0 pt] (d1) at ($(am)+2*(up)+(right)$) {};
\node [inner sep=0 pt] (d2) at ($(am)+2*(up)+3*(right)$) {};
\draw [double distance=1.5pt] (d1) to ($(d1)+1.25*(down)$);
\draw [double distance=1.5pt] (d2) to ($(d2)+1.25*(down)$);
\node (d1texte) at ($(d1)+.3*(up)$) {\large $x$};
\node (d2texte) at ($(d2)+.3*(up)$) {\large $z$};
\draw[rounded corners,fill=blue!30] ($(a1)+.25*(right)+(up)$) -- ($(a1)+1.25*(right)+1.75*(up)$) -- ($(a1)+1.25*(right)+.25*(up)$) -- ($(a1)+.25*(right)+(up)$);
\draw[ultra thick] ($(a1)+1.25*(right)+(up)$) -- ($(a1)+2.25*(right)+(up)$);
\draw[rounded corners,fill=blue!30] ($(a1)+2.25*(right)+.25*(up)$) rectangle ($(a1)+3.75*(right)+1.75*(up)$);
\draw [ultra thick] ($(am)+1.75*(right)$) -- ($(am)+2.25*(right)$);
\draw[rounded corners,fill=blue!30] ($(a1)+4.25*(right)+.25*(up)$) rectangle ($(a1)+5.75*(right)+1.75*(up)$);
\node (E_x) at ($(am)+(right)$) {\large $T_{x}$};
\node (E_z) at ($(am)+3*(right)$) {\large $T_{z}$};

\draw[rounded corners,fill=gray!20,dotted] (b1) rectangle (b2);
\node (bm) at (10,1) {};
\node (Mj) at (10.2,1.95) {\huge $\Dc$};
\node [inner sep=0 pt] (j) at ($(bm)+2*(up)+(left)$) {};
\draw [double distance=1.5pt] (j) to ($(j)+1.25*(down)$);
\node (jtexte) at ($(j)+.3*(up)+.5*(right)$) {$q \in \{\infty,0,1,2, \dots, d-1 \}$};
\node [inner sep=0 pt] (o) at ($(bm)+2*(right)$) {};
\draw [double distance=1.5pt] ($(o)+.32*(left)$) to ($(o)+(right)$);
\node (otexte) at ($(o)+1.3*(right)$) {$c$};
\draw [rounded corners, fill=blue!30] ($(b1)+.25*(up)+.25*(right)$) rectangle ($(b1)+1.75*(up)+1.75*(right)$);
\draw [ultra thick] ($(b1)+(up)+1.75*(right)$) -- ($(b1)+(up)+2.75*(right)$);

\node (m1) at ($(b1)+(up)+3.75*(right)$) {};
\draw[rounded corners,fill=blue!30] (m1) -- ($(m1)+(left)+.75*(up)$) -- ($(m1)+(left)+.75*(down)$) -- (m1);
\node (m2) at ($(b1)+1.75*(up)+2.75*(right)$) {};
\draw[rounded corners,fill=blue!30] (m2) -- ($(m2)+(right)+.75*(down)$) -- ($(m2)+1.5*(down)$) -- (m2);
\draw [thick,domain=70:110] plot ({11.17+cos(\x)}, {.12+sin(\x)});
\draw [thick,->] (11.2,1) -- (10.9,1.4) ;
\node (dj) at ($(b1)+(up)+(right)$) {\large $T_q$};

\draw[ultra thick] ($(a3)+.25*(left)$) to ($(b3)+.25*(right)$); 
\draw[densely dotted, red] ($.5*(a2)+.5*(b1)+3*(up)$) to ($.5*(a2)+.5*(b1)+2*(down)$);
\node (psi) at ($.5*(a2)+.5*(b1)+.5*(up)$) {\large $j$};

\end{tikzpicture}}
    \caption{The Torpedo Game in a prepare-and-measure (top) \textit{vs} transformational scenario (bottom). The empirical models $e_{x,z,q}$ in either scenario are the same both quantumly and classically. Hence classical and quantum strategies for any $(2,1)_d$ information  retrieval  task  can  be  equivalently  expressed in prepare-and-measure or transformational form.} \label{fig:PAM_Transform}
\end{figure}
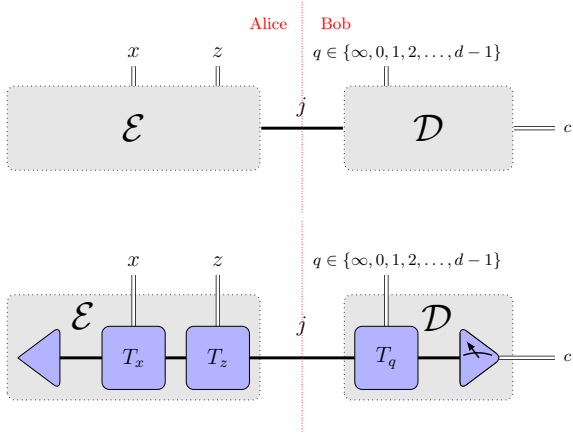

Though we focus here on the prepare-and-measure version of the Torpedo game, note that it can be equivalently expressed in a sequential scenario with fixed preparation and measurement (see \refig{PAM_Transform}). In Appendix \ref{sec:Transformational} we provide the explicit quantum and classical strategies for this purely transformational version of the Torpedo Game as well as the equivalence between prepare-and-measure and sequential protocols. 

\begin{proposition}
\label{pro:ssc}
For $d=2$ and $d=3$, strong sequential contextuality with respect to bounded memory is necessary and sufficient to win the Torpedo Game deterministically.
\end{proposition}

\begin{proof}
Suppose a bounded-memory hidden variable model realises an empirical model that wins the Torpedo Game deterministically.
Input-question combinations $(x,z,q)$ label the contexts.
Recall that the wining relation is $\omega_q(x,z)$,
and the winning condition for the torpedo game is
\begin{align}
p_e(c \not\in {\omega}_q(x,z) \mid x,z,q ) = 0. \label{eq:pam_sequential_context}
\end{align}

Using notations introduced in \refsec{emp_behaviours}, the hidden-variable model must specify probability vectors $\left\{\bm{\lambda}_{x,z}\right\}_{x,z \in \Z_d}$ and positive vectors $\left\{ \pmb{v}_q^c \right\}_{c \in \Z^d, q \in Q}$ such that $\sum_{c \in \Z_d} \pmb{v}_q^c = \pmb{1}$ for each $q \in Q$, and
\begin{align}
    \pmb{v}_q^{c} \cdot \bm{\lambda}_{x,z} &= 0\, ,
    \label{eq:linear_eq_prop}
\end{align}
for all $x,z \in \Z_d, \forall q \in Q$ and $c \not\in \omega_q(x,z)$.

As mentioned in \refsec{classical_strat}, it suffices to consider deterministic strategies. \refequ{linear_eq_prop} reduces to a set of binary linear equations (36 equations for $d=3$) that any sequentially noncontextual realisation must jointly satisfy. 

This cannot be possible since it would provide a perfect classical strategy for the $d=2$ and $d=3$ Torpedo Games, violating the optimal bounds \refequ{bounds} that were obtained by exhaustive search.
On the other hand, it is always possible to obtain a contextual realisation, by taking context-wise solutions to \refequ{linear_eq_prop}: e.g.\ where the choice of $\pmb{\lambda}_{x,z}$ is not only a function of $x$ and $z$, but also of $q$.

It can further be observed that if any fraction of an empirical model $e$ can be described noncontextually, i.e.\ $\NCF(e) = p >0$, then with an average probability at least $p$ the empirical model $e$ fails in the Torpedo Game.
Therefore, to win the Torpedo Game deterministically requires strong contextuality.

\end{proof}

An explicit noncontextual memory-bounded hidden variable model that fails to fully realise the empirical predictions but that satisfies the maximum of 33 out of 36 constraints from \refequ{linear_eq_prop} for $d=3$ is the following.
Measurement and state vectors are given in terms of the computational basis vectors, where $\pmb{\mathrm{f}}_k$ denotes the $k$th computational basis vector (zero-indexed) in the vector space $\mathbb{Z}_2^d$ over $\mathbb{R}$.
The measurement vectors are
\begin{align*}
\pmb{v}_q^c = \begin{cases} \pmb{\mathrm{f}}_{2c \oplus 2},& \quad q=\infty\\
\pmb{\mathrm{f}}_{c\oplus 2},& \quad q=0,1\\
\pmb{\mathrm{f}}_{2c\oplus 1},& \quad q=2\end{cases}
\end{align*}
The state vectors are
\begin{align*}
& \bm{\lambda}_{0,0}=\bm{\lambda}_{0,1}=\bm{\lambda}_{1,1}=\pmb{f}_0 \, , \\
& \bm{\lambda}_{1,0}=\bm{\lambda}_{0,2}=\bm{\lambda}_{2,2}=\pmb{f}_1 \, , \\
& \bm{\lambda}_{2,0}=\bm{\lambda}_{2,1}=\bm{\lambda}_{1,2}=\pmb{f}_2 \, .
\end{align*}
This corresponds to the strategy depicted in \refig{cl_d3}.

We also obtain the following more general result, of which Proposition \ref{pro:ssc} is a special case.
\begin{theorem}
\label{thm:sc}
Given any information retrieval task and strategy with empirical behaviour $e$,
\begin{equation*}
    \varepsilon \geq \textsf{\upshape NCF}(e) \, \nu \,
\end{equation*}
where $\varepsilon$ is the probability of failure, averaged over inputs and questions, $\NCF(e)$ is the bounded-memory noncontextual fraction of $e$ with memory size $d$ fixed by the scenario, and $\nu := 1- \theta^C$ measures the hardness of the task, $\theta^C$ being the classical value.
\end{theorem}

\begin{proof}
The empirical behaviour can be decomposed as:
\begin{equation*}
    e = \NCF(e) e^{\text{NC}} + \CF(e) e'
\end{equation*}
where $e'$ is necessarily strongly contextual. From this convex decomposition, we obtain that the probability of success using the empirical model $e$ reads:
\begin{equation*}
    p_{S,e} = \NCF(e) p_{S,e^{\text{NC}}} + \CF(e) p_{S,e'}
\end{equation*}
where $p_{S,e^{\text{NC}}}$ and $p_{S,e'}$ are the average probabilities associated with empirical models $e^{\text{NC}}$ and $e'$ respectively. At best, $e'$ wins with probability $1$ and thus:
\begin{align*}
    p_{S,e} & \leq \NCF(e) p_{S,e^{\text{NC}}} + \CF(e) \\
    \varepsilon & \geq \NCF(e) \varepsilon_{e^{\text{NC}}} 
\end{align*}
where $\varepsilon_{e^{\text{NC}}} = 1 - p_{S,e^{\text{NC}}}$ is the average probability of failure associated with $e^{\text{NC}}$.
Since the latter is noncontextual, we know that the minimum probability of failure is $\nu = 1 - \theta^C$, where $\theta^C$ is the classical value of the game.
Then $\varepsilon_{e^{\text{NC}}} \leq \nu$, from which we obtain the desired inequality:
\begin{equation*}
    \varepsilon \geq \NCF(e) \nu
\end{equation*}

\end{proof}

This provides a quantifiable relationship between quantum advantage and sequential contextuality.
Inequalities of this {form} are also known to arise for a variety of other informational tasks that admit quantum advantage,
with hardness measures and notions of non-classicality adapted to the particular task \cite{abramsky2017contextual,mansfield2018quantum,linde}.

    \section{Discussion}
        We have formalised a class of information retrieval tasks in communication scenarios, of which the much-studied problem of (quantum) random access coding is a special case. We showed that quantum-over-classical advantage is explained by quantum contextuality.
We have identified a distinct information retrieval task that we have presented as the Torpedo Game,
which admits a greater quantum-over-classical advantage than the comparable QRAC for qutrits by exploiting Wigner negativity. Remarkably, the qutrit torpedo strategy is maximally contextual, meaning that no fraction of it can be explained by an underlying noncontextual model.

To obtain quantum perfect strategies for the Torpedo Game we have derived a prepare-and-measure scenario for which quantum mechanics exhibits logically paradoxical behaviour (with respect to noncontextual hidden variable assumptions).
More generally, we have identified this as a characteristic that quantifies quantum advantage for any bounded-memory information retrieval task.

In the specific case of random access coding,
some works have imposed obliviousness constraints as part of the task as opposed to bounded-memory.
These restrict what information the receiver can be allowed to infer about the input string.
Whereas preparation contextuality is known to be necessary and sufficient for quantum advantage in oblivious tasks \cite{hameedi2017communication,saha2019state},
we have shown sequential contextuality to be necessary and sufficient characteristic for bounded-memory tasks.

We briefly comment on possible generalisations of the Torpedo game.
In \refequ{PerfectClassical} we conjectured, based on an explicit proof in $5 \leq d\leq 23$, that there is a perfect classical strategy in all $d\geq5$ for the Torpedo game with standard winning conditions as in \refequ{winning_conditions}.
In order to re-instate a quantum-over-classical advantage, as we had in dimensions two and three, we may modify the Torpedo game to make it harder to win classically. Note that the following modifications have no effect on the quantum values, which remain $\theta^Q_{d\geq 3}=1$. Because the $-1$ subspace of Gross' phase point operator $A$ has dimension $\frac{d-1}{2}$ it is possible to enlarge Alice's input from $d^2$ to $\frac{d^2(d-1)}{2}$. Formally, let $0\leq \ell < \frac{d-1}{2}$, so Alice sends Bob $\ket{\psi_{x,z,\ell}}=X^xZ^z \left(\ket{\ell+1}+\ket{-(\ell+1)}\right){/\sqrt{2}}$, instead of just $\ket{\psi_{x,z,\ell=0}}:=\ket{\psi_{x,z}}$ as before. The modification changes a single relation from $w_\infty (x,z) =  \{ a \in \mathbb{Z}_d \mid a \neq x \} $ to
\begin{align*}
	w_\infty(x,z,\ell)	&=\{a\in\mathbb{Z}_d|a \in \{x+\ell+1,x-\ell-1\}\},
\end{align*}
whereas the remaining conditions persist i.e, $w_q(x,z,\ell)=w_q(x,z)$ in \refequ{winning_conditions} for $q \in \{0,1,\ldots,d-1\}$. It seems reasonable that such a game, with more restrictive winning conditions, should be harder to win. Indeed, we were unable to find any perfect classical strategy by sampling, although we cannot rule out its existence since we were unable to exhaustively check all classical strategies. More generally, we have motivated how our perfect quantum strategies for this information retrieval task arise from a remarkable geometric feature of maximally negative states (c.f.~\refequ{KeyFact}), and we expect that this insight can be further mined for quantum advantage in future work.

    \section{Acknowledgements}
        The authors wish to give particular thanks to Anna Pappa for stimulating discussions that motivated the present research; to Léo Colisson, Dominik Leichtle and Nathan Shettell for first proposing the $d=3$ classical strategy that is here proved optimal; and to Miguel Navascués for sharing an example of a numerical implementation of the NPA hierarchy that allowed us to bound the quantum value of the Torpedo game for $d=2$.

Funding is gratefully acknowledged from the Irish Research Council, the French Ministry for Europe and Foreign Affairs, and the French Ministry for Higher-level Education, Research and Innovation under the PHC Ulysses Programme 2019.
M.H.\ is supported by a Royal Society–Science Foundation Ireland University Research Fellowship.
S.M.\ was supported by the Bpifrance project RISQ, and for most of the period during which this work was carried out was based at the Paris Centre for Quantum Computing and the Institut de Recherche en Informatique Fondamentale, University of Paris, where he wishes to thank Iordanis Kerenidis.

\paragraph*{Contributions} All authors contributed equally to this work. 

\vfill

    \bibliography{Torpedo}      
    
    \clearpage
    \newpage
    
    \appendix
    \section{The Transformational Version of the Torpedo Game}
    \label{sec:Transformational}
    \subsection{Transformations in Hidden-Variable Models}\label{sec:transformation}

The focus in the main text is on prepare-and-measure scenarios,
but it can also be useful to have hidden-variable representations of transformations.
A transformation $T$ is represented by a family of probability distributions
$\{ p_\tau(\cdot \mid \lambda_i, T) \}_{\lambda_i}$ over the hidden-variable space,
one for each `initial' hidden-variable $\lambda_i \in \Z_d$.
In simplified vector space notation a transformation is simply represented by a left-stochastic
matrix $\Tc_T$.
Note in particular that for transformations sequential noncontextuality requires that, e.g.,
\[ \Tc_{T_3 \circ T_2 \circ T_1} = \Tc_{T_3} \, \Tc_{T_2} \, \Tc_{T_1} \, . \]
Here we see explicitly the structural assumption that sequential composition of operations must be preserved in the hidden-variable description.
This echoes how parallel composition of measurements must be preserved in nonlocal hidden variable models.

\subsection{Equivalence between Prepare-and-Measure and Transformational Versions}

Any prepare-and-measure strategy for a $(2,1)_d$ information retrieval task (e.g., the Torpedo Game) can equivalently be re-expressed in a scenario with fixed state preparation and measurement, and with the classical inputs labelling transformations only.
This is depicted in \refig{PAM_Transform}.
The equivalence may be useful for experimental implementations, and also makes connections with other transformation-based protocols considered in \cite{zukowski2014temporal,dunjko2016quantum,clementi2017classical,mansfield2018quantum,henaut2018tsirelson}.

\begin{proposition}
\label{pro:equivalence}
Classical and quantum strategies for any $(2,1)_d$ information retrieval task can be equivalently expressed in prepare-and-measure or transformational form.
\end{proposition}

\begin{proof}
Since the initial preparation is fixed in the sequential version, it is trivial that the encoding step can always be re-expressed as a stochastic map $\mathcal{E}:\Z_d \times \Z_d \rightarrow \Z_d$ (or quantumly $\mathcal{E}:\Z_d \times \Z_d \rightarrow \mathbb{C}^d$) as in the prepare-and-measure version.
Conversely, in the classical case a strategy for the prepare-and-measure version can be expressed as a strategy for the transformational version by setting $T_x$ to always output $x$ and taking for $T_z$ the encoding map $p_\Ec(\cdot|\cdot,z)$ from the classical prepare-and-measure version.
In the quantum case $T_x$ outputs $\ket{x}$, and $T_z$ is simply taken to be a $Z$ measurement subsequently composed with the encoding map $\mathcal{E}$.
In the hidden-variable description we can fix an arbitrary basis vector, say $\pmb{\mathrm{f}}_0$, and in the quantum case an arbitrary preparation, say $\ket{0}$.

Similarly, for the decoding step it is trivial that the transformational version can always be expressed as a map in the form of the prepare-and-measure version.
For the converse, in the classical case it suffices to take for $T_q$ the stochastic decoding map $p_\Dc ( \cdot | \cdot, q)$ from the classical prepare-and-measure version, with fixed measurement given by the identity map (or in the hidden-variable description with measurement simply specified by the basis vectors).
In the quantum case, the converse follows from the observation that any projection-valued measurement can be expressed as a unitary transformation followed by a fixed measurement in the $Z$ basis.
\end{proof}

\subsection{Optimal Classical Strategy for the Transformational Version}

Note that the following analysis will hold if we consider a global transformation $\mathcal{T}_{x,z}$ instead of two transformations $\mathcal{T}_{x}$ and $\mathcal{T}_{z}$ for Alice. A perfect strategy for the Torpedo Game requires that for all $x,z \in \Z_d$, $q \in Q$ and $c \not\in \omega_q(x,z)$ that
\begin{equation}
    \Tc_q  \Tc_z  \Tc_x \ \pmb{\mathrm{f}}_0 \cdot \pmb{\mathrm{f}}_{c} = 0
    \label{eq:sequential_context}
\end{equation}

For $d=2$ it was possible to perform a brute-force search over all possible deterministic left stochastic transformations in order to check how many of the linear equations in \refequ{sequential_context} can be jointly satisfied.
As expected, at most $9$ out of $12$ equations in \refequ{sequential_context} may be jointly satisfied, matching the classical bound of \refequ{bounds}.

For $d=3$, we were unable to perform the brute-force calculation due to the size of the search space.
However the classical bound of \refequ{bounds} found by means of our grid partitioning method implies that at most $33$ out of $36$ equations in \refequ{sequential_context} may be jointly satisfied.
A solution that attains the classical value of $\frac{11}{12}$, \ie that satisfies jointly $33$ of the $36$ equations from \refequ{sequential_context}, using reversible gates only, is the following:
\begin{equation}
\begin{array}{llll}
    \Tc_{x=0} = \mathbb{I} & \Tc_{x=1} = \mathbb{I} & \Tc_{x=2} = \oplus 1 & \\
    \Tc_{z=0} = \mathbb{I} & \Tc_{z=1} = \oplus 2 & \Tc_{z=2} = \oplus 1 & \\
    \Tc_{q=\infty} = \mathbb{I} & \Tc_{q=0} = \oplus 1 & \Tc_{q=1} = \oplus 2 & \Tc_{q=2}  = \oplus 1
\end{array}
    \label{eq:classical_reversible_gates}
\end{equation}
Reversibility ensures that the strategy does not incur a simulation cost in terms of Landauer erasure, of the kind considered in \cite{henaut2018tsirelson}.
This strategy can also be implemented by states, transformations and measurements that are non-negatively represented in the discrete Wigner function, taking the stabilizer state $\ket{0}$ as initial state and representing the above permutation transformations in the obvious way. Thus the classical bound is saturated by a non-negative quantum strategy.

\subsection{Perfect Quantum Strategy for the Torpedo Game}

To explicitly establish a perfect quantum strategy in transformational form for the Torpedo Game, we re-establish the key fact \refequ{KeyFact} in the transformational setting.
Our proof uses the matrix elements of $A_{x,z}$ combined with the Clifford gates that map the computational basis to each of the additional measurement bases.
For this we use the symplectic representation of the  Clifford group.
The expressions below hold for odd prime $d$ (in the odd prime power case $d=p^n$ one should replace $\mathbb{Z}_d$ with $\mathbb{F}_d$).
Clifford group elements are written as $C=D_{x,z}U_F$ \cite{appleby2005symmetric} where
\begin{align*}
F=\left(\begin{array}{ll}
\alpha & \beta \\
\gamma & \epsilon
\end{array}\right)
\end{align*}
is an element of the symplectic group $\textrm{SL}(2,\mathbb{Z}_d)$ (entries of $F$ are in $\mathbb{Z}_d$ and $\textrm{det}\, F= 1 \bmod d$), and
\begin{align*}
U_{F}=\left\{\begin{array}{ll}
\frac{1}{\sqrt{d}} \sum_{j, k=0}^{d-1} \omega^{2^{-1}\beta^{-1}\left(\alpha k^{2}-2 j k+\epsilon j^{2}\right)}|j\rangle\langle k| & \beta \neq 0 \\
\sum_{k=0}^{d-1} \omega^{2^{-1}\alpha \gamma k^{2}}|\alpha k\rangle\langle k| & \beta=0
\end{array}\right.
\end{align*}
The matrix representation \cite{wootters1987wigner}
of a phase point operator is \begin{align}
\left(A_{x,z}\right)_{j, k}=\delta_{2 x, j+k} \omega^{z(j-k)} \label{eq:AxzMatrix}
\end{align} and so $\bra{k} A_{x,z} \ket{k}=\delta_{k,x}$ is the likelihood of getting outcome $k$ in a computational basis measurement of $A_{x,z}$.
The Clifford unitaries $\left\{U_\infty,U_0,\dots,U_{d-1}\right\}$ that map $Z=D_{0,1}$ to $\left\{D_{0,1},D_{1,0},\dots, D_{1,d-1} \right\}$ are
\begin{align}
    \{U_\infty,U_0,& \dots,U_{d-1} \} \nonumber \\
    &= \{\mathbb{I},HS^0,\dots,HS^{d-1}\} \nonumber \\
    &=
    \bigl\{
    U_{\left(\begin{smallmatrix} 1 & 0 \\ 0 & 1 \end{smallmatrix}\right)},
    U_{\left(\begin{smallmatrix} 0 & -1  \\ 1 & 0 \end{smallmatrix}\right)},
    \dots,
    U_{\left(\begin{smallmatrix} d-1 & -1 \\ 1 & 0 \end{smallmatrix}\right)}
    \bigr\}, 
\label{eq:CliffordUnitaries}
\end{align}
where $H$ and $S$ are the qudit versions of the Hadamard and Phase gate respectively. 
Using \refequ{AxzMatrix}, and the fact that $U_F A_{x,z} U_F^\dag=A_{x^\prime,z^\prime}$ where $\left(\begin{smallmatrix}x^\prime \\ z^\prime \end{smallmatrix}\right)=F\left(\begin{smallmatrix}x \\ z \end{smallmatrix}\right)$ \cite{gibbons2004discrete,gross2006hudson}, it is straightforward to verify that
\begin{align*}
\begin{split}
\bra{k} U_\infty A_{x,z} U_\infty^\dag \ket{k}&=\delta_{k,x}   \\ 
\bra{k} U_0 A_{x,z} U_0^\dag \ket{k}&=\delta_{k,-z} \\ 
& \vdots  \\
\bra{k} U_{d-1} A_{x,z} U_{d-1}^\dag \ket{k}&=\delta_{k,(d-1)x-z}\,.
\end{split}
\end{align*}
For odd prime-power $d\geq 3$, the $-1$ eigenspace of $A_{x,z}$ has rank $(d-1)/2$. We can abuse notation slightly by referring to the normalized projector onto this eigenspace as $\ketbra{\psi_{x,z}}{\psi_{x,z}}$.
The final step is to realise that $\ketbra{\psi_{x,z}}{\psi_{x,z}}=\frac{1}{d-1}\left(\mathbb{I}-A_{x,z}\right)$ so that by linearity, and in agreement with \refequ{KeyFact} earlier,

\begin{widetext}
\begin{align*}
\Tr\left(\ketbra{\psi_{x,z}}{\psi_{x,z}}\Pi_{\infty}^{k}\right)=\bra{k}U_\infty  \frac{1}{d-1}\left(\mathbb{I}-A_{x,z}\right) U_\infty^\dag \ket{k}&=\frac{1}{d-1}(1-\delta_{k,x}) \nonumber \\
\Tr\left(\ketbra{\psi_{x,z}}{\psi_{x,z}}\Pi_{0}^{k}\right)=\bra{k}U_0  \frac{1}{d-1}\left(\mathbb{I}-A_{x,z}\right) U_0^\dag \ket{k}&=\frac{1}{d-1}(1-\delta_{k,-z}) \\
& \vdots \nonumber \\
\Tr\left(\ketbra{\psi_{x,z}}{\psi_{x,z}}\Pi_{d-1}^{k}\right)=\bra{k}U_{d-1}  \frac{1}{d-1}\left(\mathbb{I}-A_{x,z}\right) U_{d-1}^\dag \ket{k}&=\frac{1}{d-1}(1-\delta_{k,(d-1)x-z}) . \nonumber
\end{align*}
\end{widetext}

Any state in the $-1$ eigenspace of $A_{x,z}$ wins the Torpedo Game with unit probability, but for concreteness we choose the state \refequ{PerfectState}.\\
\paragraph*{Circuit Version of the Optimal Quantum Strategy.}
As observed in \refprop{equivalence}, any quantum strategy for the prepare-and-measure version of the Torpedo Game admits an equivalent strategy for the transformational version.
An optimal quantum strategy in sequential operational form takes as fixed preparation $\ket{\psi_{0,0}}$ and as fixed measurement $Z$.
The transformations controlled by $x$, $z$, and $q$ are $X^x$, $Z^z$, and $U_q$, respectively, where the unitaries $U_q$ are those defined in \refequ{CliffordUnitaries}.

\begin{figure}[htbp]
	\centering
	\includegraphics[width=.7\linewidth,trim={0 1cm 0 0}]{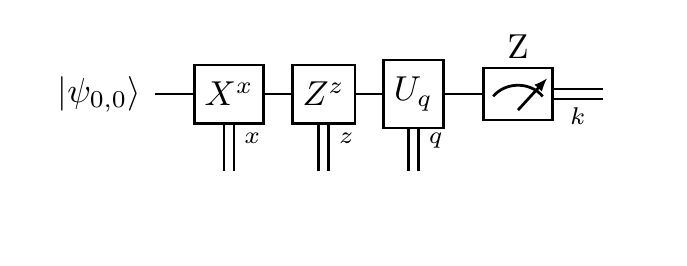}
	\caption{ \label{fig:TorpedoSequential} A perfect strategy in sequential operational form for the dimension $d$ Torpedo Game for odd power-of-prime $d$. The classically-controlled gates are appropriately defined Pauli operators or Clifford gates as in \refequ{CliffordUnitaries}.
	}
\end{figure}

    \pagebreak

\end{document}